\documentclass[journal,10pt]{IEEEtranTCOM}

\usepackage{cite}
\usepackage{graphicx}
\usepackage[cmex10]{amsmath}
\usepackage{amssymb}
\usepackage{booktabs}
\usepackage{threeparttable}
\usepackage{algorithm}
\usepackage{algpseudocode}
\usepackage{amsfonts}
\usepackage{xcolor}
\usepackage{color}

\usepackage{tabularx} %

\usepackage{caption}

\usepackage{enumitem}
\usepackage{framed}

\usepackage{dblfloatfix} %

\usepackage{soul}

\usepackage{cancel} %

\usepackage{verbatim}%

\usepackage{makecell}

\usepackage{empheq} %

\usepackage{underoverlap} %

\usepackage{hyperref}
\hypersetup{pdftex,colorlinks=true,allcolors=black}

\usepackage{calc}

\usepackage{textcomp}
\def\BibTeX{{\rm B\kern-.05em{\sc i\kern-.025em b}\kern-.08em
		T\kern-.1667em\lower.7ex\hbox{E}\kern-.125emX}}

\newtheorem{remark}{\bf  Remark}

\newtheorem{proposition}{\bf Proposition}
\newtheorem{corollary}{\bf  Corollary}

\newtheorem{lemma}{\bf  Lemma}

\newcommand{\mP}{\mathsf{P}}

\newcommand{\mj}{\mathsf{j}}

\newcommand{\myDef}{\overset{\Delta}{=}}
\newcommand{\myEqualOverset}[1]{\overset{#1}{=}}
\newcommand{\myApprOverset}[1]{\overset{#1}{\approx}}

\newcommand{\myArgMax}[1]{\underset{#1}{\mathrm{argmax}}}

\newcommand{\myFloor}[1]{\left\lfloor #1 \right\rfloor}
\newcommand{\myExp}[1]{\mathbb{E}\left\{#1\right\}}

\newcommand{\myBigO}[1]{\mathcal{O}\left(#1\right)} %

\newcommand{\myBigOsquare}[1]{\mathcal{O}\left[#1\right]} %

\newcommand{\myRangeBoth}[1]{\left[#1\right]}

\newcommand{\myComb}[2]{\left(\substack{#1\\#2}\right)}
\newcommand{\myBrcktsSqr}[1]{\left[#1\right]}
\newcommand{\myBrcktsRnd}[1]{\left(#1\right)}
\newcommand{\myBrcktsBig}[1]{\left\{#1\right\}}

\newcommand{\myNorm}[1]{\left\|#1\right\|}
\newcommand{\myDet}[1]{\mathrm{det}\left\{#1\right\}}

\newcommand{\myModulo}[2]{\left\langle#1\right\rangle_{#2}}

\newcommand{\mySpaceTwoMM}{\vspace{2mm}}

\begin{document}

\title{\huge Simultaneous Beam and User Selection for the Beamspace mmWave/THz Massive MIMO Downlink}

\author{
	Kai Wu,
	 J. Andrew Zhang,~\IEEEmembership{Senior Member,~IEEE}, %
	Xiaojing Huang,~\IEEEmembership{Senior Member,~IEEE},\\
	Y. Jay Guo,~\IEEEmembership{Fellow,~IEEE}, and Lajos
	Hanzo,~\IEEEmembership{Life Fellow, IEEE}

\thanks{{This work is partially supported by the Australian Research Council under the Discovery Project Grant DP210101411.} 
}

\thanks{L. Hanzo would like to acknowledge the financial support of the 
	Engineering and Physical Sciences Research Council projects EP/W016605/1 
	and EP/X01228X/1 as well as of the European Research Council's Advanced 
	Fellow Grant QuantCom (Grant No. 789028)}

	\thanks{K. Wu, J. A. Zhang, X. Huang and Y. J. Guo are with the Global Big Data Technologies Centre, University of Technology Sydney, Sydney, NSW 2007, Australia (e-mail: kai.wu@uts.edu.au; andrew.zhang@uts.edu.au; xiaojing.huang@uts.edu.au;   jay.guo@uts.edu.au).}%
	
	\thanks{L. Hanzo is with the Department of Electronics and Computer Science, University of Southampton, Southampton SO17 1BJ, UK (e-mail:
		lh@ecs.soton.ac.uk).}
}

\maketitle

\begin{abstract}	
	Beamspace millimeter-wave (mmWave) and terahertz (THz) massive MIMO constitute attractive schemes for next-generation communications, given their abundant bandwidth and high throughput.
	However, their
	user and beam selection problem has not been efficiently addressed yet. 
	Inspired by this challenge, we develop
	low-complexity solutions explicitly. 
	In contrast to the zero forcing in the prior art, we 
	introduce the dirty paper coding (DPC) into the joint user and beam selection problem. {We unveil the compelling properties of the DPC sum rate in beamspace massive MIMO, showing its monotonic evolution against the number of users and beams selected. We then exploit its beneficial properties for substantially simplifying the joint user and beam selection problem. Furthermore, we develop a set of algorithms striking unique trade-offs for solving the simplified problem, facilitating simultaneous user and beam selection based on partial beamspace channels for the first time. 
		Additionally, we derive the sum rate bound of the algorithms and analyze their complexity.} Our simulation results validate the effectiveness of the proposed design and analysis, confirming their superiority over prior solutions. 
\end{abstract}


\section{Introduction}\label{sec: introduction}

Millimeter-wave (mmWave) and terahertz (THz) massive multiple-input and multiple-output (MIMO) communications are expected to become a reality in the next-generation era \cite{6G_vision_2020network,6G_XiaohuYou2021towards,6G_enable2020Access}. Hybrid antenna arrays constitute a key ingredient of mmWave/Thz massive MIMO systems. Among other popular types, the analog multi-beam antenna array concept has attracted extensive attention due to its appealingly low power consumption in passive beamforming \cite{Lajos_mmWaveCom_survey2018,Lens_lens5GlowPower_2018Mag,KaiWu_WIPT_LAA}. Such antenna arrays can be realized with the aid of circuit-type beamforming networks or quasi-optical lens \cite{Lens_PathDivision_TCOM2016,Jay_Butler6G_overview,Jay_lens6G_overview}. Either way, the analog beamforming part often relies on multiple beams. The so-called beamspace massive MIMO exploits these beams, since communications take place in a space spanned by the beams \cite{Beamspace_sayeed2013beamspaceMIMO}.

To compensate for the severe propagation loss experienced in the mmWave and THz frequency bands, massive MIMO antenna arrays generally have a large dimension, leading to a large number of beams \cite{Lajos_BmSlctPrecdng_JSTSP2018,Lajos_beamspaceWidebandChannelEstimation_TSP2019,Lens_beamspaceChannelEstimation_TCM2018}. Restricted by factors such as power consumption and the form factor, the number of radio frequency (RF) chains is usually much smaller than that of the beams. 
Hence, how to select a subset of beams for best performance is a key problem in beamspace massive MIMO.

The seminal beam selection work \cite{Beamspace_sayeed2013beamspaceMIMO} opts for the specific beams bearing the strongest channel power for each user. The method is appealingly simple but may suffer from inter-user interference. An interference-aware method is proposed in \cite{Lens_beamSelection_XinyuGao2016}, which 
assigns a beam to a user, if only that user achieves the strongest channel power in the beam. We refer to it as \textit{the strongest beam} of the user for convenience. For users sharing strongest beams with others, an incremental beam search is developed in \cite{Lens_beamSelection_XinyuGao2016}, iteratively searching for the specific beam to maximize the sum rate in combination with previously selected beams. The zero-forcing transmit precoding (ZF-TPC) is used for deriving the sum rate \cite{Lens_beamSelection_XinyuGao2016}.

By employing ZF, incremental and also decremental beam search (removing a beam in each iteration to minimize sum rate loss) methods are developed in \cite{Lens_beamSelec_TCOM2015_Masouros}. A two-stage beam selection scheme is designed in \cite{Lens_BeamSelect2stages_2019}. Using ZF, a sum rate lower bound is derived and used for formulating a convex problem for the initial beam selection. Then, an iterative algorithm is designed for refining the selection results. 
As a further development, the dirty paper coding (DPC) is first introduced to the beam selection problem in \cite{Lens_beamSelectionDPC_2018Pal}, where a 
decremental beam selection method is developed. The complexity of the above methods is further reduced in \cite{Lens_BeamSelect_coplexityReduc2021_WCL} using innovative matrix manipulations.

While the contributions reviewed above mainly consider the beam selection problem, the importance of user selection for beamspace massive MIMO is highlighted in 
\cite{Lens_JonitUserBeamSelection_2020}.
Indeed, user selection (scheduling) has become indispensable in contemporary mobile communications networks, e.g., LTE and 5G \cite{book_ahmadi2019_5G}. 
The authors of \cite{Lens_JonitUserBeamSelection_2020} first find the best matching user subset for a candidate beam subset and then search for the user subset for maximizing the ZF sum rate. 
However, given $ M $ RF chains, $ K $ users and $ N $ beams, the authors assume that the optimal number of users/beams may range from one to $ M $, 
hence the total number of combinations of user/beam subsets 
may become excessive, given by $ \sum_{m=1}^{M} \myComb{K}{m}\times \myComb{N}{m} $. This may become unrealistic for real-time communications.

{Hence, we are inspired to substantially reduce the complexity of user and beam selection for beamspace massive MIMO systems. To do so, we introduce DPC into the joint selection problem, going way beyond the prior art of using ZF. 
We first unveil the appealing properties of the DPC-based sum rate and then exploit them for drastically simplifying
the joint user and beam selection problem. Three algorithms striking different design trade-offs are developed for solving this challenging problem. Their performance vs. complexity is also analyzed. Our main contributions are summarized as follows. 

\begin{enumerate}
	\item We show the monotonic evolution of the DPC sum rate vs. the number of selected users and beams. We prove that selecting the same number of users and beams as that of the RF chains is a sufficient condition for maximizing the DPC sum rate. This substantially reduces the number of subsets of users/beams to be probed, as compared to the prior art \cite{Lens_JonitUserBeamSelection_2020}. Using the example mentioned earlier, 
	this number is reduced from 
	$ \sum_{m=1}^{M} \myComb{K}{m}\times \myComb{N}{m} $ \cite{Lens_JonitUserBeamSelection_2020} to $ \myComb{K}{M}\times \myComb{N}{M} $.
	
	\item  We develop three attractive algorithms striking different design trade-offs for solving the simplified user and beam selection problem. The first algorithm sequentially selects users first and then beams, while the second and third algorithms select users and beams simultaneously. An incremental greedy search is employed in the algorithms. In contrast to previous user/beam selection solutions, the null projection technique is applied, in line with DPC, for iteratively maximizing the sum-rate.
	For the simultaneous user and beam selection, we arrange to use partial beamspace channels for the first time, further reducing the
	complexity imposed.

	\item  We derive the sum-rate upper bound of the proposed algorithms. We also analyze their computational complexity, in comparison to previous methods. By design, the proposed user and beam selection algorithms have an even lower complexity than the sole stand-alone beam selection method (also using DPC) of \cite{Lens_beamSelectionDPC_2018Pal}, yet their sum rate performance is similar. 
\end{enumerate}
In a nutshell, we boldly and explicitly compare our our novel contributions to the related prior art in Table \ref{tab: novel contributions}. }

\begin{table}[!t]\footnotesize
	\captionof{table}{Highlighting novel contributions in contrast to prior arts}
	\vspace{-3mm}
	\begin{center}
		\begin{tabular}{m{3.5cm}|m{0.5cm}|m{0.5cm}|l|l|l}
			\hline
			Features &
			\textbf{Our work} &
			\cite{Beamspace_sayeed2013beamspaceMIMO,Lens_beamSelection_XinyuGao2016,Lens_beamSelec_TCOM2015_Masouros,Lens_BeamSelect2stages_2019}
			& \cite{Lens_beamSelectionDPC_2018Pal}
			& \cite{Lens_BeamSelect_coplexityReduc2021_WCL}
			& \cite{Lens_JonitUserBeamSelection_2020}
			\\
			\hline
			
			Beam selection
			& $ \checkmark $
			& $ \checkmark $
			& $ \checkmark $
			& $ \checkmark $
			& $ \checkmark $
			\\				
			\hline
			
			User selection  
			& $ \checkmark $
			& 
			& 
			& 
			& $ \checkmark $
			\\				
			\hline
			
			ZF
			&
			& $ \checkmark $ 
			& 
			& $ \checkmark $ 
			& $ \checkmark $ 
			\\				
			\hline
			
			DPC
			&  $ \checkmark $ 
			& 
			& $ \checkmark $ 
			& $ \checkmark $ 
			& 
			\\				
			\hline

			Fixed number of beams or users to be selected
			& $ \checkmark $ 
			& $ \checkmark $ 
			& $ \checkmark $ 
			& $ \checkmark $ 
			& 
			\\				
			\hline
			
			Sum rate monotonicity against the number of selected users or beams 
			& $ \checkmark $ 
			&  
			& 
			& 
			& 
			\\				
			\hline

			Sum rate upper bound for the joint user and beam selection
			& $ \checkmark $ 
			& 
			& 
			& 
			& 
			\\				
			\hline
			
			Efficient, simultaneous user and beam selection
			& $ \checkmark $ 
			& 
			& 
			& 
			& 
			\\				
			\hline

		\end{tabular}
		\vspace{-3mm}
	\end{center}
	\label{tab: novel contributions}
\end{table}

Section \ref{sec: signal model} establishes the signal model and formulates the joint user and beam selection problem. Section \ref{sec: simplify selection problem} analyzes the DPC sum rate properties mentioned above and simplifies the joint selection problem. 
Section \ref{sec: proposed selection algorithms} develops three algorithms for solving the simplified problem, followed by performance-vs-complexity analysis in Section \ref{sec: performance complexity analysis}. Our simulation results are provided in Section \ref{sec: simulation results}, and we conclude the paper in Section \ref{sec: conclusion}. 

Throughout the paper, the following notations/symbols are used, unless specified otherwise. Bold upper-case letters are used for matrices; bold lower-case for vectors; calligraphic upper-case letters, e.g., $ \mathcal{I} $, for sets. $ (\cdot)^{\mathrm{T}} $ denotes transpose; $ (\cdot)^{\mathrm{H}} $ for conjugate transpose. 
$ \mathrm{diag}\{\cdots\} $ generates a diagonal matrix with the enclosed values put on the diagonal; $ \mathbf{I}_x $ denotes an $ x $-dimensional identity matrix; $ \mathrm{det}\{\} $ represents the matrix determinant; $ \mathbb{E}\{\} $ is the expectation; $ \mathbb{B} $ denotes the set of Boolean numbers; $ \mathbb{C} $ is the set of complex numbers; $ \mathbb{R} $ is the set of real numbers; 
$ \myBrcktsSqr{\mathbf{x}}_n $ denotes the $ n $-th entry of a vector $ \mathbf{x} $; $ \myBrcktsSqr{\mathbf{X}}_{a,b} $ is the $ (a,b) $-th entry of the matrix $ \mathbf{X} $; $ \myBrcktsSqr{\mathbf{X}}_{a,:} $ is the $ a $-th row; $ \myBrcktsSqr{\mathbf{X}}_{:,b} $ is the $ b $-th column; $ \myBrcktsSqr{\mathbf{X}}_{a,\mathcal{I}} $ is the $ a $-th row with the column entries indexed by elements in the set $ \mathcal{I} $; $ \myBrcktsSqr{\mathbf{X}}_{\mathcal{I},:} $ represents the rows indexed by elements in $ \mathcal{I} $, while $ \myBrcktsSqr{\mathbf{X}}_{:,\mathcal{I}} $ represents the columns.

\begin{figure*}[!t]
	\centering
	\includegraphics{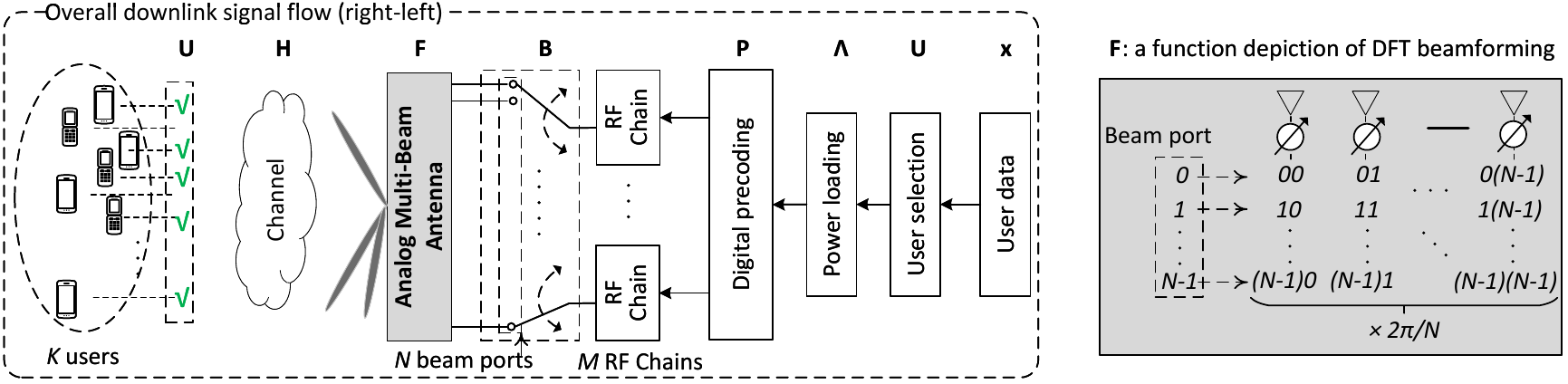}
	\caption{The schematic diagram of the overall beamspace MIMO downlink focused on in this work. The signal flow is from right to left, as complied with the signal model given in (\ref{eq: y bf = ...x + z}). 
		Schematically, we assume each RF chain can access all beam ports through the connected switch. 
		The analog multi-beam antenna performs the full-dimensional DFT beamforming whose functional depiction is given on the right.
		One can think of DFT beamforming as  
		using the columns of a DFT matrix for the beamforming weights in a uniform linear array.
		This shall not be confused with practical implementations of multi-beam antennas \cite{Jay_Butler6G_overview,Jay_lens6G_overview}.}\label{fig: system diagram}
\end{figure*}

\section{Signal Model and Problem Formulation} \label{sec: signal model}

As illustrated in Fig. \ref{fig: system diagram}, we consider the beamspace mmWave/THz massive MIMO downlink relying on a multi-antenna transmitter and single-antenna users. {The extension to multi-antenna users or the mix of single- and multi-antenna users can be straightforward, as will be illustrated in Remark \ref{rmk: multiple antenna at users}.} {The transmitter is equipped with an analog multi-beam antenna array and multiple RF chains to support different spatial streams. The analog array performs fixed discrete Fourier transform (DFT) based beamforming \cite{Kai_Expeditious2019TWC}. It
	can be implemented either using quasi-optical lenses or Butler matrix-based beamforming network circuits \cite{Jay_Butler6G_overview,Jay_lens6G_overview}.}
 {A functional depiction of the DFT based beamformer is given in Fig. \ref{fig: system diagram}. The $ n $-th $ (n=0,1,\cdots,N-1) $ beam port is associated with a so-called DFT beam that may be deemed equivalent to setting the beamforming weights of an $ N $-antenna uniform linear array as the $ n $-th column of the $ N\times N $ DFT matrix.}

In massive MIMO communications, 
the number $ N $ of DFT beams can be much larger than the number $ M $ of RF chains. Moreover, we consider the typical scenario of $ K(\gg M) $ users. 
{As in previous seminal beam/user selection contributions \cite{Lens_beamSelection_XinyuGao2016, Lens_beamSelec_TCOM2015_Masouros, Lens_BeamSelect2stages_2019, Lens_beamSelectionDPC_2018Pal, Lens_BeamSelect_coplexityReduc2021_WCL,Lens_JonitUserBeamSelection_2020}, each RF chain selects one beam at most.}
\textit{Thus, our task is to select $ B(\le M) $ out of $ N $ beams to serve $ U(\le B) $ out of $ K $ users so that the sum rate of the downlink communications described above is maximized.}

Let $ \mathbf{H} $ denote the $ K\times N $ channel matrix.
The $ k $-th row, as denoted by $ \mathbf{h}_k\in \mathbb{C}^{1\times N} $, is the channel vector of the $ k $-th user upon employing the widely used Saleh-Valenzuela model for mmWave/THz channels \cite{Lens_xinyuGao_compressiveChannelEstimation,Beamspace_gao2016fast_THz_channeltracking}, $ \mathbf{h}_k $ can be written as $ \mathbf{h}_k = \sum_{l=0}^{L_k} \alpha_{kl} \mathbf{a}^{\mathrm{T}}(u_{kl}) $, where $ L_k $ denotes the number of paths, $ \alpha_{kl} $ is the path gain, and $  \mathbf{a}(u_{kl})\in\mathbb{C}^{N\times 1} $ is the steering vector. The $ n $-th entry of $ \mathbf{a}(u_{kl}) $ can be formulated as $ \myBrcktsSqr{\mathbf{a}(u_{kl})}_n = e^{\mj n u_{kl}} $. 
Here, $ u_{kl} $ is the spatial frequency of the $ l $-th path. We have $ u_{kl}=\pi\sin\theta_{kl}\in\myRangeBoth{-{\pi},{\pi}} $ in a uniform linear array having half-wavelength antenna spacing, where $ \theta_{kl}\in\myRangeBoth{-\frac{\pi}{2},\frac{\pi}{2}} $ denotes the spatial angle. 

{As illustrated in Fig. \ref{fig: system diagram}, the DFT beamforming performed by the analog multi-beam antenna array is equivalent to taking the DFT of signals over the $ N $ beam ports. Let $ \mathbf{F} $
denote the $ N $-dimensional DFT matrix.
The channel matrix $ \mathbf{H} $ is then turned into $ \tilde{\mathbf{H}} = \mathbf{H}\mathbf{F} $, termed as the so-called beamspace channel matrix.} The $ (a,b) $-th entry of the matrix is $\myBrcktsSqr{\mathbf{F}}_{a,b}=\frac{1}{N}e^{-\mj \frac{2\pi ab }{N}}~(a,b=0,1,\cdots,N-1) $. The coefficient $ 1/N $ is used for power normalization. 
Let $ \mathbf{x}\in \mathbb{C}^{K\times 1} $ denote the communication signal vector of all users. The signals received by the selected users can be stacked into the following vector, 
\begin{align} \label{eq: y bf = ...x + z}
	\mathbf{y} = \mathbf{U} \tilde{\mathbf{H}} \mathbf{B} \mathbf{P} \mathbf{\Lambda} \mathbf{U} \mathbf{x} + \mathbf{z}, ~\mathrm{s.t.}~\tilde{\mathbf{H}}\myDef \mathbf{H} \mathbf{F},
\end{align}
where $ \mathbf{z} $ is a complex white Gaussian noise vector, and the component matrices, as marked in Fig. \ref{fig: system diagram}, are described below.
\begin{enumerate}
	\item $ \mathbf{U} \in\mathbb{B}^{U\times K} $ is a user selection matrix, selecting $ U $ out of $ K $ users. It is a sub-matrix of $ \mathbf{I}_{K} $, hence we have $ \mathbf{U}\subset \mathbf{I}_{K} $;
	
	\item $ \tilde{\mathbf{H}}\in\mathbb{C}^{K\times N} $ is known at the transmitter for beam and user selection, as {in previous work \cite{Lens_beamSelection_XinyuGao2016, Lens_beamSelec_TCOM2015_Masouros, Lens_BeamSelect2stages_2019, Lens_beamSelectionDPC_2018Pal, Lens_BeamSelect_coplexityReduc2021_WCL,Lens_JonitUserBeamSelection_2020}};
	
	\item $ \mathbf{B}  \in\mathbb{B}^{N\times B} $ is a beam selection matrix, selecting $ B(\le M) $ out of $ N $ DFT beams. It is a sub-matrix of $ \mathbf{I}_N $, indicating that a single beam is selected for each user and conversely each selected user is assigned a single beam;
	
	\item $ \mathbf{P}\in\mathbb{C}^{B\times U} $ is a TPC matrix used for suppressing the inter-user interference;
	
	\item $ \mathbf{\Lambda} = \mathrm{diag}\{\cdots,\sqrt{\lambda_u},\cdots\}\in\mathbb{R}^{U\times U} $ is a diagonal matrix adjusting the transmission power for each user.

\end{enumerate}
Now, the beam and user selection problem can be modeled as:
\begin{align} \label{eq: max S}
	& \max_{U,B,\mathbf{U},\mathbf{B},\mathbf{\Lambda}} \mathcal{S}\myDef \log_2{   \det \myBrcktsBig{ \frac{1}{\sigma_z^2}\myExp{\mathbf{y}\mathbf{y}^{\mathrm{H}}}} } \\
	 \mathrm{s.t.}~&\|\mathbf{B} \mathbf{P} \mathbf{\Lambda} \mathbf{U} \mathbf{x}\|_2^2\le \mP,\nonumber\\
	& \mathbf{U}(\in\mathbb{B}^{U\times K}) \subset \mathbf{I}_K,~1\le U\le B \le M \ll K,\nonumber\\
	& \mathbf{B}  (\in\mathbb{B}^{N\times B} ) \subset \mathbf{I}_N,~1\le B\le M \ll N,\nonumber
\end{align}
where $ \myDet{\cdot} $ denotes the determinant, the objective function is the sum rate, and $ \mP $ denotes the overall transmission power. {Upon substituting (\ref{eq: y bf = ...x + z}) into (\ref{eq: max S}), we can get the following typical sum rate expression
	\begin{align}
		&	\mathcal{S} = \log_2{   \det \myBrcktsBig{ \mathbf{I}_U + \rho \breve{\mathbf{H}} \mathbf{R}_{xx} \breve{\mathbf{H}}^{\mathrm{H}}  }
		},~~\mathrm{s.t.}~ \rho = \mP/\sigma_z^2;\nonumber\\
		&~~~~~~~~\breve{\mathbf{H}} = \mathbf{U} \tilde{\mathbf{H}} \mathbf{B};~ \mathbf{R}_{xx} = \frac{1}{\mP}\mathbb{E}\myBrcktsBig{\mathbf{P} \mathbf{\Lambda} \mathbf{U} \mathbf{x} (\mathbf{P} \mathbf{\Lambda} \mathbf{U} \mathbf{x})^{\mathrm{H}}}, \nonumber
	\end{align}
	where $ \mathbf{I}_U $ denotes an $ U $-dimensional identity matrix, $ \breve{\mathbf{H}} $ is the equivalent beamspace channel (factoring in both the user and beam selection), and $ \mathbf{R}_{xx} $ is the auto-correlation matrix of the communication signals over selected users. Note that the sum rate expression above is written based on the definition of MIMO channel capacity, namely the mutual information between the transmitted and received signal vectors \cite{book_mimoOFDMmatlab}. The inter-user interference is considered through the determinant operation. 
}

Optimally solving the problem is difficult, not only due to its non-convex nature but also owing to the enormous size of the feasible region. Moreover, the specific expression of $ \mathcal{S} $ varies with the choice of TPC methods. {Based on the authoritative literature \cite{Beamspace_sayeed2013beamspaceMIMO,Lens_beamSelection_XinyuGao2016,Lens_beamSelec_TCOM2015_Masouros,Lens_BeamSelect2stages_2019,Lens_BeamSelect_coplexityReduc2021_WCL,Lens_JonitUserBeamSelection_2020}, we have learned that the ZF sum rate generally has a complex non-convex expression, which is challenging to deal with in optimization. To circumvent this challenge, authors have often turned to maximizing the lower/upper bounds of the ZF sum rate. Since the tightness of the bounds cannot be guaranteed, maximizing the bounds requires iterative optimization. Each iteration tends to involve further greedy searches. These difficulties in dealing with ZF-based optimization for beam/user selection inspire us to harness and characterize a different precoding scheme.}

{Explicitly, DPC is selected by us, as it is known to have a simple sum rate expression \cite{book_mimoOFDMmatlab}, eliminating the need for lower/upper bounds of the sum rate to be maximized. 
	Even better, we will prove shortly that the DPC sum rate of the beamspace massive MIMO considered has nice properties enabling us to substantially simplify the user and beam selection problem.
	At the time of writing, DPC has only been applied for beam selection in beamspace massive MIMO in \cite{Lens_beamSelectionDPC_2018Pal}, but the above compelling properties have not been unveiled before. 
}

\begin{remark} \label{rmk: multiple antenna at users}
	{Multiple antennas can be available at a user. Since a single beam is selected for each user at the transmitter, it is beneficial to have a multi-antenna user receiver steer a focused beam towards the specific direction corresponding to the transmitted beam. This can be realized using low-complexity phase shifters at the user end, hence substantially reducing the power consumption. 
		In this case, the signal model established above is still applicable. We only have to modify the path gain $ \alpha_{kl} $ (for the $ l $-th path of the $ k $-th user) by factoring in the receiver beamforming performed at a user.} 
\end{remark}

\section{Simplifying the Beam and User Selection}
\label{sec: simplify selection problem}

In this section, we first reveal the relationship between the sum rate achieved by DPC and the numbers of selected beams ($ B $) and user ($ U $). Then we exploit this relationship for substantially simplifying the beam- and user-selection problem.

DPC is performed based on the QR decomposition of the channel matrix \cite{book_mimoOFDMmatlab}. The composite channel matrix of our system is $ \mathbf{U} \tilde{\mathbf{H}} \mathbf{B} $ based on (\ref{eq: y bf = ...x + z}). Following the QR decomposition, we have 
\begin{align} \label{eq: U H B = QR}
	(\mathbf{U} \tilde{\mathbf{H}} \mathbf{B})^{\mathrm{H}} = \mathbf{Q}\mathbf{R},~\mathrm{s.t.}~\mathbf{Q}^{\mathrm{H}}\mathbf{Q}=\mathbf{I}_U,~
\end{align}
where $ \mathbf{R} $ is a $ U\times U $ upper triangular matrix. 
Performing DPC is equivalent to setting $ \mathbf{P} $ as 
\begin{align} \label{eq: P = QRD}
	\mathbf{P} = \mathbf{Q}(\mathbf{R}^{\mathrm{H}})^{-1}\mathbf{D},~
	\mathrm{s.t.}~[\mathbf{D}]_{u,u} = r_u\myDef [\mathbf{R}^{\mathrm{H}}]_{u,u}, 
\end{align}
where $ \mathbf{D} $ is a $ U\times U $ diagonal matrix.
In QR decomposition, $ r_u\ne0~(\forall u) $ can be made positive. Moreover, $ r_u=0 $ can be readily excluded in precoding. Hence, \textit{$ r_u>0~(\forall u) $ is assumed by default hereafter.}

Substituting (\ref{eq: U H B = QR}) and (\ref{eq: P = QRD}) into (\ref{eq: y bf = ...x + z}) yields $ \mathbf{y}=\mathbf{D}\mathbf{\Lambda}\mathbf{U}\mathbf{x} $. Then, the sum rate $ \mathcal{S} $ given in (\ref{eq: max S}) becomes
\begin{align} \label{eq: S DPC}
	&\mathcal{S}_{\mathrm{DPC}} = \sum_{u=0}^{U-1} \log_2\myBrcktsRnd{1 + \frac{r_u^2 \lambda_u \sigma_s^2}{\sigma_z^2}}\myEqualOverset{\substack{ \sigma_s^2=1\\ \sigma_z^2 =1}}\sum_{u=0}^{U-1} \log_2\myBrcktsRnd{1 + r_u^2 \lambda_u },
\end{align}
where $ \sqrt{\lambda_u} $ is the $ u $-th diagonal entry of $ \mathbf{\Lambda} $, $ \sigma_s^2 $ denotes the power of the $ k $-th $ (\forall k\in[0,K-1]) $ entry of $ \mathbf{x} $ (information signal vector), and $ \sigma_z^2  $ is the noise variance.
\textit{For simplicity, we take $  \sigma_s^2=1 $, as is often the case in practice, and use the power loading factor $ \lambda_u $ for controlling the actual transmitted power for user $ u $. We also take $ \sigma_z^2=1  $ hereafter (including the simulations) and adjust $ \mP $ to set the signal-to-noise ratio (SNR).}

 Given any combination of feasible $ U,B,\mathbf{U} $ and $ \mathbf{B} $, maximizing $ \mathcal{S}_{\mathrm{DPC}} $ under the power constraint given in (\ref{eq: max S}) leads to the popular water-filling solution \cite{book_mimoOFDMmatlab} 
\begin{align} \label{eq: lambda u water filling}
	\lambda_u = \myBrcktsSqr{ \beta - \frac{1}{r_u} }_+,~\mathrm{s.t.}~\sum_{u=0}^{U-1} \lambda_u = \mP,%
\end{align}
where $ [x]_+=\max \{x,0\} $. 
Substituting (\ref{eq: lambda u water filling}) into (\ref{eq: S DPC}) yields
\begin{align} \label{eq: S DPC +}
	\mathcal{S}_{\mathrm{DPC}} = \sum_{u=0}^{U-1} \myBrcktsSqr{\log_2\myBrcktsRnd{ \beta r_u } }_+.
\end{align}
The sum rate $ \mathcal{S}_{\mathrm{DPC}} $ is a function of $ U $ and $ B $ (as implied in $ r_u $). 
Thus, if we can ascertain the monotonicity of the function vs. these parameters, we can then exploit it for determining the optimal values of $ U $ and $ B $ first. This motivates the following exploration.

\textit{For simplicity, we consider the asymptotic case relying on an adequate $ \mP $, so that the $ [\cdot]_+ $ in (\ref{eq: lambda u water filling}) and (\ref{eq: S DPC +}) can be removed.} We note that this asymptotic condition may not be necessary for the results to be derived; however, without it, the analytical illustration may become tediously complicated. 

Consider the general case of $ U=U_1~(\forall U_1<B) $ and denote the corresponding DPC sum rate by $ \mathcal{S}_1 $. Let furthermore $ \beta=\beta_1 $ be the upper limit used in the water filling; see (\ref{eq: lambda u water filling}). Based on (\ref{eq: S DPC +}) and given an adequate $ \mP $, we have $ \mathcal{S}_1 = \sum_{u=0}^{U_1-1} {\log_2(\beta_1r_u)} $. 
Let us now include an additional user $ u=U_1+1 $. 
Given an adequate but fixed power budget $ \mP $, we have to deduct power from the previous $ U_1 $ users and assign it to the new user. 
Under the water filling power allocation, the upper limit $ \beta_1 $ has to decrease, say by $ \delta $, to cater for an extra user. Based on (\ref{eq: lambda u water filling}), the power taken from the previous $ U_1 $ users is given by $ U_1\delta $, while the power assigned to the new user is $ \beta_1-\delta-\frac{1}{\gamma_{U_1+1}} $. Equating the two amounts gives $ \gamma_{U_1+1}=\frac{1}{\beta_1-(U_1+1)\delta} $. Modifying $ \mathcal{S}_1 $ given above, the sum rate including the new user can be formulated by $ \mathcal{S}_2 = \sum_{u=0}^{U_1-1} {\log_2((\beta_1-\delta)r_u)} + \log_2(\frac{(\beta_1-\delta)}{\beta_1-(U_1+1)\delta}) $. The difference between the sum rates is 
\begin{align}
	\mathcal{S}_2 - \mathcal{S}_1 = \log_2\myBrcktsRnd{\frac{(\beta_1-\delta)^{U_1+1}}{\beta_1 -(U_1+1)\delta}} - \log_2\myBrcktsRnd{\beta_1^{U_1}}\nonumber.
\end{align}
The first derivative of the difference with respect to (w.r.t.) $ \delta $ is always positive. Moreover, we have $ \lim\limits_{\delta\rightarrow 0_+} \mathcal{S}_2 - \mathcal{S}_1 = 0 $. Thus, $ \mathcal{S}_2 - \mathcal{S}_1>0 $ is ensured for any $ \delta>0 $. The above discussions are summarized below.

\mySpaceTwoMM

\begin{lemma} \label{lm: sum rate U, user number}
	\textit{Given an adequate $ \mP $, the DPC sum rate $ \mathcal{S}_{\mathrm{DPC}} $ increases with the number $ U $ of selected users.}
\end{lemma}

\mySpaceTwoMM

The relationship between $ \mathcal{S}_{\mathrm{DPC}} $ and the number of selected beams $ B $ is more subtle. To reveal it, we first provide a useful result whose proof is given in Appendix \ref{app: proof of lemma prod(r_u) vs B}. 

\mySpaceTwoMM

\begin{lemma} \label{lm: prod(r_u) vs B}
	\textit{The absolute product of the diagonal entries of $ \mathbf{R} $, i.e., $ {\Pi_{u=0}^{U-1}r_u} $, is non-decreasing with $ B $, where $ \mathbf{R} $ is the upper triangular matrix obtained in (\ref{eq: U H B = QR}) and $ B $ is the number of selected beams.}
\end{lemma}

\mySpaceTwoMM

Let us denote the DPC sum rates at $ B=B_1~(\forall B_1<M) $ and $ B=B_1+1 $ as $ \mathcal{S}_1 =\sum_{u=0}^{U-1} \log_2\myBrcktsRnd{\beta r_u} $ and $ \mathcal{S}_2 =\sum_{u=0}^{U-1} \log_2\myBrcktsRnd{(\beta+\epsilon) \tilde{r}_u} $, respectively. Note that the change of $ B $ leads to the changes in $ {r}_u $ and $ \beta $, as reflected by $ \tilde{r}_u $ and $(\beta+\epsilon)  $. Based on the water filling in (\ref{eq: lambda u water filling}), we have $ \sum_{u=0}^{U-1} \beta-\frac{1}{r_u} = \mP $ and $ \sum_{u=0}^{U-1} \beta+\epsilon-\frac{1}{\tilde{r}_u} = \mP $, which lead to $ \beta = \frac{1}{U}\myBrcktsRnd{\mP+\sum_{u=0}^{U-1} \frac{1}{r_u} } $ and $ \beta+\epsilon = \frac{1}{U}\myBrcktsRnd{\mP+\sum_{u=0}^{U-1} \frac{1}{\tilde{r}_u} } $, respectively. Substituting them back to the sum rates and after some manipulations (with details suppressed), we arrive at
\begin{align}
	& \mathcal{S}_1 = 
\log_2\myBrcktsRnd{ 
	\myBrcktsRnd{\frac{\mP+\sum_{u=0}^{U-1} \frac{1}{r_u}}{U} }^U 
	{ (\Pi_{u=0}^{U-1}r_u) }   }; \nonumber
\end{align}
\begin{align}
	& \mathcal{S}_2 =  \log_2\myBrcktsRnd{ 
			\myBrcktsRnd{\frac{\mP+\sum_{u=0}^{U-1} \frac{1}{\tilde{r}_u}}{U} }^U 
			{ (\Pi_{u=0}^{U-1}\tilde{r}_u) }   }.\nonumber
\end{align}
Note that $ \tilde{r}_u~(\forall u) $ is obtained at $ B_1+1 $, which is surely larger than $ B_1 $ for $ r_u $. 
Upon applying Lemma \ref{lm: prod(r_u) vs B}, we have $ \Pi_{u=0}^{U-1}\tilde{r}_u\ge \Pi_{u=0}^{U-1}{r}_u $. 
In high-SNR conditions, i.e., for $ r_u,\tilde{r}_u\gg 1~(\forall u) $, we can observe that the first term within the two logarithmic functions is approximately the same. The above observation supports the following statement.

\mySpaceTwoMM

\begin{lemma}\label{lm: sum rate vs B}
	\textit{Given high SNRs, implying $ r_u\gg 1~(\forall u) $, the DPC sum rate $ \mathcal{S}_{\mathrm{DPC}} $ is a non-decreasing function of the number of selected beams, i.e., $ B $.}
\end{lemma}

\mySpaceTwoMM

Ensured by Lemmas \ref{lm: sum rate U, user number} and \ref{lm: sum rate vs B}, 
the DPC sum rate is a non-decreasing function of $ U $ and $ B $; hence, 
a more important and interesting insight is provided below. 

\mySpaceTwoMM

\begin{proposition} \label{pp: sum rate vs B U}
	\textit{For the beamspace mmWave/THz massive MIMO downlink considered, a sufficient condition for maximizing the DPC sum rate is selecting $ U=M $ users and $ B=M $ beams, where $ M $ is the number of RF chains.} 
\end{proposition}

\mySpaceTwoMM

The results obtained above can be used for simplifying the beam and user selection problem modeled in (\ref{eq: max S}). In particular, the objective function in (\ref{eq: max S}) can now be replaced with the DPC sum rate $ \mathcal{S}_{\mathrm{DPC}} $ given in (\ref{eq: S DPC +}). Enabled by Proposition \ref{pp: sum rate vs B U}, we can now take $ U = M $ and $ B = M $. Consequently, the optimization variables are reduced to $ \mathbf{U} $ and $ \mathbf{B} $ solely, since the optimal $ \mathbf{\Lambda} $ can be obtained based on the water filling given in (\ref{eq: lambda u water filling}). Upon exploiting all the above changes, the beam and user selection problem can be 
reformulated as 
\begin{align} \label{eq: max S simplified}
	& \max_{\mathbf{U},\mathbf{B}}~ \mathcal{S}_{\mathrm{DPC}}=\sum_{u=0}^{M-1} \myBrcktsSqr{\log_2\myBrcktsRnd{ \beta r_u } }_+ \\
	\mathrm{s.t.}~& \mathbf{U}(\in\mathbb{B}^{M\times K}) \subset \mathbf{I}_K,~ \mathbf{B}  (\in\mathbb{B}^{N\times M} ) \subset \mathbf{I}_N,\nonumber \\
	& r_u\myDef [\mathbf{R}^{\mathrm{H}}]_{u,u},~(\mathbf{U} \tilde{\mathbf{H}} \mathbf{B})^{\mathrm{H}} = \mathbf{Q}\mathbf{R},~\mathbf{Q}^{\mathrm{H}}\mathbf{Q}=\mathbf{I}_U,~ \nonumber\\
	& \lambda_u = \myBrcktsSqr{ \beta - \frac{1}{r_u} }_+,~\sum_{u=0}^{U-1} \lambda_u = \mP. \nonumber   %
\end{align}
We {develop} algorithms below for solving this substantially simplified problem, as compared with the original problem modeled in (\ref{eq: max S}).

\section{Proposed Beam and User Selection} \label{sec: proposed selection algorithms}

Due to its non-convex nature, solving Problem (\ref{eq: max S simplified}) is still difficult. Basically, all previous {related studies \cite{Lens_beamSelection_XinyuGao2016, Lens_beamSelec_TCOM2015_Masouros, Lens_BeamSelect2stages_2019, Lens_beamSelectionDPC_2018Pal, Lens_BeamSelect_coplexityReduc2021_WCL,Lens_JonitUserBeamSelection_2020}} have turned to sub-optimal greedy search. 
We proceed by developing such sub-optimal solutions as well. The majority of the previous contributions only consider beam selection, i.e., solely searching for $ \mathbf{B} $ with $ K=M $ and hence $ \mathbf{U}=\mathbf{I} $. This suggests that, if we can first determine the optimal $ M $ users out of the $ K $ candidates, the remaining problem then solely involves the beam selection.
This leads to the first algorithm presented in Section \ref{subsec: sequential user and beam selection}.

\subsection{Sequential User and Beam Selection} \label{subsec: sequential user and beam selection}
In this algorithm, we first select the $ M $ users that maximize the DPC sum rate under the full beamspace channel (i.e., with all beams involved). We then select the $ M $ beams for maximizing the DPC sum rate for the pre-determined users, as in the sole beam selection problem. 
Given the beamspace channel matrix $ \tilde{\mathbf{H}} $, the first user we select only has to have the strongest channel power, since no interference emanating from other users needs to be considered yet. 
Thus, the first user is simply selected via 
\begin{align} \label{eq: m star first}
	u^{\star}:~\myArgMax{u\in[0,K-1]}~\myNorm{\myBrcktsSqr{\tilde{\mathbf{H}} }_{u,:} }_2^2;~~~\mathcal{U}^{\star} = \{ u^{\star} \}, 
\end{align}
where $ \mathcal{U}^{\star} $ collects the optimal users.

For the $ u $-th $ (u\ge 1) $ selected user, we have to maximize the channel power and meanwhile minimize the interference arriving from the users already selected. Such a user can be selected by maximizing the power of the projection of a channel vector (under test) onto the null space of the previously selected channel vectors \cite{UserSelection_ZF_Dimic2005TSP}. 
The null space can be constructed as 
\begin{align} \label{eq: N bf U star}
	\mathbf{N}_{\mathcal{U}^{\star}} = \mathbf{I}_N - \bar{\mathbf{H}}^{\mathrm{H}}\myBrcktsRnd{\bar{\mathbf{H}}\bar{\mathbf{H}}^{\mathrm{H}}  }^{-1}\bar{\mathbf{H}},~~\mathrm{s.t.}~~ \bar{\mathbf{H}} = \myBrcktsSqr{\tilde{\mathbf{H}} }_{\mathcal{U}^{\star},:},
\end{align}
where $ \myBrcktsSqr{\tilde{\mathbf{H}} }_{\mathcal{U}^{\star},:} $ denotes a sub-matrix formed by the rows of $ \tilde{\mathbf{H}} $ indexed by the elements of the set $ \mathcal{U}^{\star} $. 
Upon employing $ \mathbf{N}_{\mathcal{U}^{\star}} $, the next optimal user can be identified as
\begin{align} \label{eq: m star next (M-1)}
	& u^{\star}:~\myArgMax{u\in[0,K-1]\setminus \mathcal{U}^{\star}}~\myNorm{ \myBrcktsSqr{\tilde{\mathbf{H}} }_{u,:}\mathbf{N}_{\mathcal{U}^{\star}} }_2^2;~~~\mathcal{U}^{\star} = \mathcal{U}^{\star}\cup u^{\star},
\end{align}
where $ (\cdot)\setminus (\cdot) $ returns the set difference. \textit{Performing the above maximization up to $ (M-1) $ times, the $ M $ users that maximize the DPC sum rate under the full beamspace channel are identified.} 
Thus, for the sequential beam selection, we take the following $ \mathbf{U} $ in (\ref{eq: max S simplified}), i.e.,
\begin{align} \label{eq: U=U*=[I]_U*}
\mathbf{U} = 	\mathbf{U}^{\star}=\myBrcktsSqr{\mathbf{I}_K}_{\mathcal{U}^{\star},:}.
\end{align}

Finding $ \mathbf{B} $ in (\ref{eq: max S simplified}) can be done by the popular incremental or decremental greedy search \cite{Lens_beamSelec_TCOM2015_Masouros}.
However, in contrast to {all previous treatises \cite{Beamspace_sayeed2013beamspaceMIMO,Lens_beamSelection_XinyuGao2016, Lens_beamSelec_TCOM2015_Masouros, Lens_BeamSelect2stages_2019, Lens_beamSelectionDPC_2018Pal, Lens_BeamSelect_coplexityReduc2021_WCL,Lens_JonitUserBeamSelection_2020}}, we continue using the above null space projection technique for identifying the optimal beams. Similar to (\ref{eq: m star first}), the first optimal beam is identified as 
\begin{align} \label{eq: b star first}
	b^{\star}:~\myArgMax{b\in[0,N-1]}~\myNorm{\myBrcktsSqr{\mathbf{U}^{\star}\tilde{\mathbf{H}}}_{:,b} }_2^2;~~~\mathcal{B}^{\star} = \{b^{\star}\}, 
\end{align}
where $ \mathcal{B}^{\star} $ collects the optimal beams. Then the $ b $-th $ (b\ge 1) $ beam is iteratively identified as 
\begin{align} \label{eq: b star next (M-1)}
	b^{\star}:~\myArgMax{b\in[0,N-1]\setminus \mathcal{B}^{\star}}~\myNorm{\myBrcktsSqr{\mathbf{U}^{\star}\tilde{\mathbf{H}}}_{:,b}^{\mathrm{H}}  \mathbf{N}_{\mathcal{B}^{\star}} }_2^2;~~\mathcal{B}^{\star} = \mathcal{B}^{\star}\cup b^{\star}.
\end{align}
Note that $ \mathbf{N}_{\mathcal{B}^{\star}} $ is the following null space constructed before solving the above problem:
\begin{align} \label{eq: N bf B*}
	\mathbf{N}_{\mathcal{B}^{\star}} = \mathbf{I}_M - \breve{\mathbf{H}}\myBrcktsRnd{\breve{\mathbf{H}}^{\mathrm{H}}\breve{\mathbf{H}}  }^{-1}\breve{\mathbf{H}}^{\mathrm{H}}~,\mathrm{s.t.}~ \breve{\mathbf{H}} = \myBrcktsSqr{\mathbf{U}^{\star}\tilde{\mathbf{H}}}_{:,\mathcal{B}^{\star}},
\end{align}
where $ \myBrcktsSqr{\mathbf{U}^{\star}\tilde{\mathbf{H}}}_{:,\mathcal{B}^{\star}} $ denotes a sub-matrix formed by the columns of the enclosed matrix indexed by the elements in the set $ \mathcal{B}^{\star} $. 
Upon iteratively solving (\ref{eq: b star next (M-1)})
for up to $ (M-1) $ times, the $ M $ beams that maximize the DPC sum rate with the pre-selected $ M $ users are obtained. The sequential user and beam selection procedure illustrated above is 
summarized in Algorithm \ref{alg: user first and then beams}. 

\begin{algorithm}[!t]
	\small
	\caption{\small Sequential User and Beam Selection}
	\label{alg: user first and then beams}
	\textit{Input:} Beamspace channel matrix $ \tilde{\mathbf{H}} $; see (\ref{eq: y bf = ...x + z}); $ M $ (RF chain number); $ K $ (user number); $ N $ (beam number)
	
	\textit{Output:} $ \mathcal{U}^{\star} $ and $ \mathcal{B}^{\star} $
	
	\vspace{2mm}
	\textit{User selection}
	\begin{enumerate}
		\item Solve (\ref{eq: m star first}) for the initial $ \mathcal{U}^{\star} $;
		
		\item For $ m=1:M-1 $, perform:

		\begin{enumerate}
			
			\item Construct the null space $ \mathbf{N}_{\mathcal{U}^{\star}} $ given in (\ref{eq: N bf U star});	
			\item For each $ u\in[0,K-1]\setminus \mathcal{U}^{\star} $, perform:

			\begin{enumerate}
				
				\item Calculate the projection in the objective function of (\ref{eq: m star next (M-1)});
			\end{enumerate}

			\item Solve (\ref{eq: m star next (M-1)}) to select a new user, and update $ \mathcal{U}^{\star} $;
		
		\end{enumerate}
\end{enumerate}

		\vspace{2mm}
		\noindent \textit{Beam selection}
\begin{enumerate}	
	\item  Set $ \mathbf{U} = 	\mathbf{U}^{\star}=\myBrcktsSqr{\mathbf{I}_K}_{\mathcal{U}^{\star},:} $;
		
		\item Solve (\ref{eq: b star first}) for the initial $ \mathcal{B}^{\star} $; 
		
\item For $ m=1:M-1 $, perform:

\begin{enumerate}
	\item Construct the null space $ \mathbf{N}_{\mathcal{B}^{\star}} $ given in (\ref{eq: N bf B*});
	\item For each $ b\in[0,N-1]\setminus \mathcal{B}^{\star} $, perform:
	
	\begin{enumerate}

		\item Calculate the projection in the objective function of (\ref{eq: b star next (M-1)});
	\end{enumerate}

	\item Solve (\ref{eq: b star next (M-1)}) to select the next beam, and update $ \mathcal{B}^{\star} $;
	
\end{enumerate}
\end{enumerate}

\end{algorithm}

We note that in the joint user and beam selection, the $ M $ users that are optimal under the full channel with all beams involved may not be optimal under the partial channel associated with the selected beams.
	A key reason for this is that we only use $ M $ beams, one for each user, for communications. When using all beams for user selection, we also have to deal with extra interference that may not exist in communications. However, Algorithm \ref{alg: user first and then beams} is not without merits. It actually provides a performance upper bound for the joint user and beam selection problem, as it will be illustrated in Section \ref{sec: performance complexity analysis}. Next, we develop another algorithm, reducing the unnecessary interference we have to deal with.

\subsection{Simultaneous User and Beam Selection}\label{subsec: simultaneous user and beam selection}
Instead of selecting users first and then beams, we propose next to select them simultaneously. 
In so doing, we first provide an important result, as proved in Appendix \ref{app: prood of propostion on M threshold}. \textit{For convenience, we refer to the beam corresponding to the strongest channel power of a user as its strongest beam.} 

\mySpaceTwoMM

\begin{proposition} \label{pp: M threashold}
	\textit{Let us represent the specific event that the strongest beams of the $ M $ (out of $ K $) users maximizing the DPC sum rate have different indexes by $ \mathcal{E} $. Then, the probability of $ \mathcal{E} $
		tends to one, provided that
		\begin{align} \label{eq: M threshold}
			M\le \bar{M}\myDef \myFloor{\frac{KN}{K+N}},
		\end{align}
	where $ \myFloor{\cdot} $ represents flooring, $ K $ is the total number of users, and $ N $ is that of the beams. 
	}
\end{proposition}

\mySpaceTwoMM

Equipped with Proposition \ref{pp: M threashold}, we now proceed to develop an algorithm simultaneously selecting a user and a beam under a pair of  complementary cases: \textit{1) the condition (\ref{eq: M threshold}) is satisfied; or 2) it is not}. The first step is the same in both cases, namely identifying the user and the beam having the highest channel power:
\begin{align}\label{eq: u b first}
	\{u^{\star},b^{\star}\}:~\myArgMax{\substack{u\in[0,K-1]\\b\in[0,N-1]}}~\myNorm{\myBrcktsSqr{\tilde{\mathbf{H}}}_{u,b}}_2^2;~
	\begin{array}{c}
		\mathcal{U}^{\star} = \{u^{\star}\}\\
		\mathcal{B}^{\star} = \{b^{\star}\}	
	\end{array},
\end{align}
where $ \tilde{\mathbf{H}} $ is the beamspace channel matrix given in (\ref{eq: y bf = ...x + z}).
The remaining $ (M-1) $ users and $ (M-1) $ beams are selected in the ensuing $ (M-1) $ iterations; each iteration has a user and also a beam selected.

\mySpaceTwoMM

\textbf{Case 1}: \textit{the condition (\ref{eq: M threshold}) is satisfied}.
To select the next user and beam, we project each candidate channel vector onto a null space spanned by the selected users and beams. Unlike in Algorithm \ref{alg: user first and then beams}, the null space is no longer constructed based on the full channel information; instead, it is constructed based on the beams already selected plus the strongest beam of the user under test. The rationale is twofold: 1) the strongest beam contributes most to the received power; 2) according to Proposition \ref{pp: M threashold}, the strongest beam of the next user to be selected has not yet been selected almost for sure, provided that the condition (\ref{eq: M threshold}) holds.

For the $ u $-th user under test, we identify its strongest beam and generate a beam index set as follows,
\begin{align} \label{eq: I cal}
	\mathcal{I} = \myBrcktsBig{\mathcal{B}^{\star},\myArgMax{b\in[0,N-1]}\myNorm{\myBrcktsSqr{\tilde{\mathbf{H}}}_{u,b} }_2^2 },
\end{align}
where $ \mathcal{B}^{\star} $ is the set of beams selected previously. 
Note that $ \mathcal{I} $ has a cardinality higher than $ \mathcal{U}^{\star} $ by one. 
Next, we construct the null space of the selected users with the beams in $ \mathcal{I} $, i.e., 
\begin{align} \label{eq: N bf U* I}
	\mathbf{N}_{\mathcal{U}^{\star}}^{\mathcal{I}} = \mathbf{I}_{|\mathcal{I}|} - \check{\mathbf{H}}^{\mathrm{H}}\myBrcktsRnd{\check{\mathbf{H}}\check{\mathbf{H}}^{\mathrm{H}}  }^{-1}\check{\mathbf{H}},~\check{\mathbf{H}} = \myBrcktsSqr{\tilde{\mathbf{H}}}_{\mathcal{U}^{\star},\mathcal{I}}. 
\end{align}
We then project the partial beamspace channel vector of the $ u $-th user in $ \mathcal{I} $ onto the above null space and calculate the projection power
\begin{align} \label{eq: Pu}
	\mathcal{P}_u = \myNorm{ \myBrcktsSqr{\tilde{\mathbf{H}}}_{u,\mathcal{I}} \mathbf{N}_{\mathcal{U}^{\star}}^{\mathcal{I}}}_2^2,~\forall u\in [0,K-1]\setminus\mathcal{U}^{\star}.
\end{align}
The user and its strongest beam, leading to the highest projection power, will be selected next, i.e., 
\begin{align}\label{eq: u b next M-1}
	& 
	\begin{array}{c}
		\mathcal{U}^{\star} = \mathcal{U}^{\star}\cup u^{\star} \\
		\mathcal{B}^{\star} = \mathcal{B}^{\star} \cup b^{\star}	
	\end{array};~
	 \mathrm{s.t.}~ \{u^{\star},b^{\star}\}:~\myArgMax{\substack{u\in[0,K-1]\setminus\mathcal{U}^{\star}}}~\mathcal{P}_u, %
\end{align}
where $ b^{\star} $ is the index of the strongest beam of user $ u^{\star} $. The simultaneous user and beam selection procedure is summarized in Algorithm \ref{alg: user and beams together}.

\mySpaceTwoMM

\textbf{Case 2}: \textit{the condition (\ref{eq: M threshold}) is NOT satisfied}. In this case, 
only probing the strongest beam of each user candidate may lead to poor sum rate performance, since more than one selected users can achieve the strongest channel power in the same beam. However, according to Proposition \ref{pp: M threashold}, we can readily state that the first $ \bar{M} $ users to be selected will not share the strongest beams almost for sure. Consequently, to select the first $ \bar{M} $ users, we can continue solving (\ref{eq: u b first}) and (\ref{eq: u b next M-1}). Then, for the remaining $ (M-\bar{M}) $ users, we have to test more than one beam for each user candidate. Since the second strongest beam can only be one of the neighbors in the vicinity of the strongest beam, we propose to test the three beams\footnote{Testing more beams can result in negligible performance difference, as will be validated by Fig. \ref{fig: sum rate vs i No.} in Section \ref{sec: simulation results}.}.

Based on $ \mathcal{I} $ given in (\ref{eq: I cal}), we construct 
the following three beam sets, as differentiated by the superscript,
\begin{align} \label{eq: I tilde i}
	& \tilde{\mathcal{I}}^{i} = \myBrcktsBig{\mathcal{B}^{\star},\myModulo{b^{\star}+i}{N} },~i=\{0,\pm 1\},\nonumber\\
	&\mathrm{s.t.}~b^{\star}:~\myArgMax{b\in[0,N-1]}\myNorm{\myBrcktsSqr{\tilde{\mathbf{H}}}_{u,b} }_2^2,
\end{align}
where $ \myModulo{x}{N} $ denotes the modulo-$ N $ of $ x $. For each beam set, we can construct a null space as done in (\ref{eq: N bf U* I}). Then, we can project the partial beamspace channel vector of the user under test onto the null space, and calculate the projection power. The above three steps can be jointly described by
\begin{align} \label{eq: P tilde u i}
	& \tilde{\mathcal{P}}_u^i = \myNorm{ \myBrcktsSqr{\tilde{\mathbf{H}}}_{u,\tilde{\mathcal{I}}^i} \mathbf{N}_{\mathcal{U}^{\star}}^{\tilde{\mathcal{I}}^i}}_2^2,~\begin{array}{l}
		i=\{0,\pm1\};\\
		\forall u\in [0,K-1]\setminus\mathcal{U}^{\star}
	\end{array} \\
	& \mathrm{s.t.}~\mathbf{N}_{\mathcal{U}^{\star}}^{\tilde{\mathcal{I}}^i} = \mathbf{I}_{|\tilde{\mathcal{I}}^i|} - \check{\mathbf{H}}^{\mathrm{H}}\myBrcktsRnd{\check{\mathbf{H}}\check{\mathbf{H}}^{\mathrm{H}}  }^{-1}\check{\mathbf{H}},~\check{\mathbf{H}} = \myBrcktsSqr{\tilde{\mathbf{H}}}_{\mathcal{U}^{\star},\tilde{\mathcal{I}}^i}. \nonumber
\end{align}
The user and beam, leading to the highest projection power, will be selected next, i.e., 
\begin{align}\label{eq: u i for u b best}
	& 
\mathcal{U}^{\star} = \mathcal{U}^{\star}\cup u^{\star};~ 
\mathcal{B}^{\star} = \mathcal{B}^{\star} \cup b^{\star}	+i^{\star}, \nonumber\\
	 \mathrm{s.t.}~& \{u^{\star},i^{\star}\}:~\myArgMax{\substack{u\in[0,K-1]\setminus\mathcal{U}^{\star}\\
			i\in\{0,\pm 1\}	
	}}~\tilde{\mathcal{P}}_u^i;~b^{\star} \text{ in (\ref{eq: I tilde i})}. %
\end{align}

\begin{algorithm}[!t]
	\small
	\caption{\small Simultaneous User and Beam Selection}
	\label{alg: user and beams together}
	\textit{Input:} Beamspace channel matrix $ \tilde{\mathbf{H}} $; see (\ref{eq: y bf = ...x + z}); $ N $ (beam number); $ K $ (user number); $ M $ (RF chain number)

	\textit{Output:} $ \mathcal{U}^{\star} $ and $ \mathcal{B}^{\star} $
	
	\begin{enumerate}
		\item Calculate $ \bar{M} $ based on Proposition \ref{pp: M threashold};
		
		\item Solve (\ref{eq: u b first}) for initial $ \mathcal{U}^{\star} $ and $ \mathcal{B}^{\star} $;
		
		\item \textit{For $ m=1:\min\{M,\bar{M}\} -1$, iteratively perform:}

		\begin{enumerate}
			\item \textit{For each $ u\in [0,K-1]\setminus \mathcal{U}^{\star} $, perform:}
			
			\begin{enumerate}
				\item Construct the beam index set $ \mathcal{I} $ as per (\ref{eq: I cal});
				
				\item Construct the null space $ \mathbf{N}_{\mathcal{U}^{\star}}^{\mathcal{I}} $ as per 
				(\ref{eq: N bf U* I});
				
				\item Calculate the projection $ \mathcal{P}_u $ given in (\ref{eq: Pu});

			\end{enumerate}
			
			\item Solve (\ref{eq: u b next M-1}) for the next user and beam; update $ \mathcal{U}^{\star} $ and $ \mathcal{B}^{\star} $;

		\end{enumerate}
		
		\vspace{1mm}
		
		\item {If $ M>\bar{M} $, proceed; otherwise, terminate.}
		
		\vspace{1mm}
		
		\item \textit{For $ m=\bar{M}:M-1 $, iteratively perform:}
		
		\begin{enumerate}
			\item \textit{For each $ u\in [0,K-1]\setminus \mathcal{U}^{\star} $ and each $ i=\{0,\pm 1\} $, perform:}
			
			\begin{enumerate}
				\item Construct the beam index set $ \tilde{\mathcal{I}}^{i}  $ as per (\ref{eq: I tilde i});

				\item Calculate the projection $ \tilde{\mathcal{P}}_u^i $ given in (\ref{eq: P tilde u i});

			\end{enumerate}
			
			\item Solve (\ref{eq: u i for u b best}) for the next user and beam; update $ \mathcal{U}^{\star} $ and $ \mathcal{B}^{\star} $;
		\end{enumerate}

	\end{enumerate}
\end{algorithm}

 Again, 
 Algorithm \ref{alg: user and beams together} summarizes the two cases for the proposed simultaneous user and beam selection. The $ \bar{M} $ derived in Proposition \ref{pp: M threashold} is used as a threshold, so that we can proceed up to Step 4) if $ M\le \bar{M} $ or proceed further if the condition does not hold.
 Next, we provide a variant of Algorithm \ref{alg: user and beams together} having a lower complexity and a slight performance loss. 
 However, the asymptotic sum rate performance  
 of the next algorithm is the same as Algorithm \ref{alg: user and beams together}, as it will be seen shortly.

\subsection{Low-Complexity Simultaneous User and Beam Selection} \label{subsec: sir based user and beam selection}
Algorithms \ref{alg: user first and then beams} and \ref{alg: user and beams together} employ null space projection to search for the optimal beams and/or users in each iteration. However, the construction of null spaces can be computationally heavy. The reason for using the null space projection is to look for the next user (beam) that suffers from the minimum interference inflicted by the previous users (beams). 
In the mmWave/THz massive MIMO system considered, the null space projection can be approximated by matrix multiplications, thanks to the following result. 

\mySpaceTwoMM

\begin{corollary}\label{col: channel orthogonal asymp}
	\it The strongest channel components of the $ M $ selected users are asymptotically orthogonal, given a sufficiently large $ K $ and $ N $. 	
\end{corollary}

\begin{IEEEproof}
	Given $ K\gg M $, Proposition \ref{pp: M threashold} suggests that the strongest beams of the $ M $ selected user are different with the probability approaching one. 
	On the other hand, the channel components underlain by the strongest beams are asymptotically orthogonal as $ N $ increases \cite[Lemma 1]{Lens_xinyuGao_compressiveChannelEstimation}.
\end{IEEEproof}

\mySpaceTwoMM

Corollary \ref{col: channel orthogonal asymp}
further suggests that the rows of $ \check{\mathbf{H}} $ can be approximately orthogonal. Thus, we can suppress the null space projection and turn to our conventional but effective metric, namely the signal-to-interference ratio (SIR). Referring to (\ref{eq: P tilde u i}), $ \myBrcktsSqr{\tilde{\mathbf{H}}}_{u,\tilde{\mathcal{I}}^i} $
is the channel vector of the user under test, and $ \check{\mathbf{H}} $ is the channel matrix of selected users. The signal power is the power in $ \myBrcktsSqr{\tilde{\mathbf{H}}}_{u,\tilde{\mathcal{I}}^i} $, while the interference power can be obtained from the inner product between $ \myBrcktsSqr{\tilde{\mathbf{H}}}_{u,\tilde{\mathcal{I}}^i} $ and $ \check{\mathbf{H}} $. 
Thus, we can define the following SIR,
\begin{align}\label{eq: SIR}
	\gamma_u^i  \approx \myNorm{ \myBrcktsSqr{\tilde{\mathbf{H}}}_{u,\tilde{\mathcal{I}}^i}  }_2^2 \Big/ \myNorm{ \myBrcktsSqr{\tilde{\mathbf{H}}}_{u,\tilde{\mathcal{I}}^i} \check{\mathbf{H}}^{\mathrm{H}} }_2^2.
\end{align}
The beam and user selection procedure using (\ref{eq: SIR}) is summarized in Algorithm \ref{alg: simul user and beams low complexity}.

\begin{algorithm}[!t]
	\small
	\caption{\small Simultaneous User and Beam Selection with Lower Complexity}
	\label{alg: simul user and beams low complexity}
	
	While sharing most of Algorithm \ref{alg: user and beams together}, 
	
	1. Replace $ \mathcal{P}_u $ in Step 3)-a)-iii) with $ \gamma_u^i~(i=0) $ in (\ref{eq: SIR}); 
 
2. Replace $ \tilde{\mathcal{P}}_u^i $ in Step 5)-a)-ii) with $ \gamma_u^i ~(i=0,\pm 1)$ in (\ref{eq: SIR}). 
\end{algorithm}

\section{Performance and Complexity Analysis}\label{sec: performance complexity analysis}

We analyze the sum rate performance first and then the computational complexity of the proposed user and beam selection algorithms.

\subsection{Performance Analysis}

\begin{corollary} \label{col: sum rate upper bound}
	\it The upper bound of the sum rate attained by Algorithms \ref{alg: user first and then beams}-\ref{alg: simul user and beams low complexity} can be expressed as
	\begin{align}
		\bar{\mathcal{S}} = \mathcal{S}_{\mathrm{DPC}}\big|_{\mathbf{U}=\mathbf{U}^{\star},\mathbf{B}=\mathbf{I}_N},
	\end{align} 
where the right-hand side is obtained by substituting the variable values in the subscript into (\ref{eq: max S simplified}), and $ \mathbf{U}^{\star} $ is given in (\ref{eq: U=U*=[I]_U*}).
\end{corollary}

\begin{IEEEproof}	
	Since $ \mathbf{B} $ is the beam selection matrix, taking $ \mathbf{B}=\mathbf{I}_N $
	means that all beams are used for calculating the sum rate, i.e., without beam selection. Note that $ \mathbf{B}=\mathbf{I}_N $ is feasible for the sum rate computation in (\ref{eq: max S simplified}), but it is infeasible practically, since it requires the same number of RF chains as that of DFT beams.

Upon applying Lemma \ref{lm: sum rate vs B},  we have that $ \mathbf{B}=\mathbf{I}_N $ is a sufficient condition for maximizing the DPC sum rate, given $ \mathbf{U}=\mathbf{U}^{\star} $. 
As given in (\ref{eq: U=U*=[I]_U*}),	
	$ \mathbf{U}^{\star} $ is constructed based on the optimal $ \mathcal{U}^{\star} $ obtained by the first part, up to Step 2c), of Algorithm \ref{alg: user first and then beams}. All three algorithms use the incremental user selection. Thus, we can state that $ \mathcal{U}^{\star} $ obtained by the first part of Algorithm \ref{alg: user first and then beams} leads to the maximum sum rate, since it is the only one employing the full channel information.  
\end{IEEEproof}

\mySpaceTwoMM

\begin{corollary} \label{col: alg 2 better K=M}
	\it 	For $ K=M $, Algorithm \ref{alg: user and beams together} achieves a sum rate upper bounded by Algorithm \ref{alg: user first and then beams}.
\end{corollary}

\begin{IEEEproof}
	The condition of $ K=M $ indicates that no user selection is necessary. In other words, Algorithms \ref{alg: user first and then beams} and \ref{alg: user and beams together} return the same set of selected users. %
	
	Both algorithms use the incremental greedy beam selection. 
	However, Algorithm \ref{alg: user first and then beams} always involves all users in each iteration, while Algorithm \ref{alg: user and beams together} does not. Moreover, according to Lemma \ref{lm: sum rate U, user number}, we know that the sum rate is an increasing function of the number of users. 
\end{IEEEproof}

\mySpaceTwoMM

\begin{remark}\label{rmk: K=M}
	\textit{For $ K=M $, the sum rate of Algorithm \ref{alg: user and beams together} can be close to that of Algorithm \ref{alg: user first and then beams}.} This is mainly due to the threshold $ \bar{M} $ derived in Proposition \ref{pp: M threashold}. This allows us to achieve approximate optimality for the first $ \bar{M} $ selected users, since their strongest channel components are nearly orthogonal, according to Corollary \ref{col: channel orthogonal asymp}. The threshold $ \bar{M} $ also allows us to probe extra beams for the remaining $ (M-\bar{M}) $ users, hence substantially reducing sum rate loss. 
\end{remark}

\begin{remark}\label{rmk: K gg M}
	{\textit{For $ K\gg M $ and a large $ N $, the sum rate performance of Algorithm \ref{alg: simul user and beams low complexity} approaches that of Algorithm \ref{alg: user and beams together}. Both algorithms
			generally achieve a higher sum rate than Algorithm \ref{alg: user first and then beams}.} 
		The condition $ K\gg M $ implies that we can almost always find $ M $ users having different strongest beams, as dictated by Proposition \ref{pp: M threashold}. Then, Corollary \ref{col: channel orthogonal asymp} indicates that these users have nearly orthogonal strongest channel components. 
		Thus, maximizing the SIR given in (\ref{eq: SIR}), as relied on by Algorithm \ref{alg: simul user and beams low complexity}, becomes approximately equivalent to maximizing the $ \ell_2 $-norm of the null space projection given in (\ref{eq: Pu}), as adopted by Algorithm \ref{alg: user and beams together}. The equivalent metrics allow these two algorithms to achieve similar sum rates. This will be further confirmed in Section \ref{sec: simulation results} through simulations.
}
\end{remark}
	
\begin{remark} \label{rmk: optimal }
	{We note that the choice of Algorithms \ref{alg: user first and then beams}-\ref{alg: simul user and beams low complexity} is not ideal for Problem (\ref{eq: max S}). Exhaustive search is known to be optimal, but has excessive complexity \cite{Lens_beamSelection_XinyuGao2016}. The incremental search adopted substantially reduces the complexity of the exhaustive search, but it is sub-optimal. 
		Intuitively, this is because we choose to employ some reasonable yet sub-optimal metrics, e.g., maximizing the null space projection in (\ref{eq: u b next M-1}), to make it more feasible to solve Problem (\ref{eq: max S}). However, remarkably, Algorithms \ref{alg: user first and then beams}-\ref{alg: simul user and beams low complexity} are asymptotically optimal in the case of $ K=M $, as $ N $ becomes sufficiently large.
		According to Remark \ref{rmk: K gg M}, a sufficiently large $ N $ suggests that the beamspace channels of the $ M $ users are orthogonal. Thus, the three algorithms, striving for finding the strongest beams having the best orthogonality, are guaranteed to return the optimal set of beams.} 
\end{remark}

\subsection{Computational Complexity} \label{subsec: compelxity}

\subsubsection{User Selection in Algorithm \ref{alg: user first and then beams}}
Step 1) has a complexity order of $ \myBigO{KN} $. 
Step 2a) has a complexity order $ \myBigO{m^2} $, which only accounts for computing $ (\bar{\mathbf{H}}\bar{\mathbf{H}}^{\mathrm{H}})^{-1} $
in the null space. The other matrix multiplication will be jointly evaluated with Step 2b-i). 
Note that the direct computation of the inverse has a complexity of $ \myBigO{m^3} $, but here we employ the iterative update of the matrix inverse \cite{UserSelection_ZF_Dimic2005TSP}. 
Based on (\ref{eq: N bf U star}) and (\ref{eq: m star next (M-1)}), the complexity of Step 2b-i) is given by $ \myBigO{N+Nm+m^2+m} $, where $ m $ is the cardinality of $ \mathcal{U}^{\star} $ in the $ m $-th iteration and the norm in (\ref{eq: b star next (M-1)}) is computed via $  \myNorm{\myBrcktsSqr{\tilde{\mathbf{H}} }_{u,:}\mathbf{N}_{\mathcal{U}^{\star}}}_2^2 = \myBrcktsSqr{\tilde{\mathbf{H}} }_{u,:}\mathbf{N}_{\mathcal{U}^{\star}}\myBrcktsSqr{\tilde{\mathbf{H}} }_{u,:}^{\mathrm{H}} $. 
According to Step 2b), Step 2b-i) is performed $ (K-m) $ times in the $ m $-th $ (m\in [1,M-1]) $ iteration. 
To sum up, the overall complexity of the user selection is
\begin{align} \label{eq: complexity of user selection in Alg 1}
	&\sum_{m=1}^{M-1} \myBigOsquare{(K-m)(N+Nm+m^2+m)} + \myBigO{m^2}\nonumber\\
	& + \myBigO{KN}
	\myApprOverset{N\gg M; K\gg M} \myBigO{KM^2N}.
\end{align}

\subsubsection{Beam Selection in Algorithm \ref{alg: user first and then beams}}
Step 2) has a complexity of $ \myBigO{MN} $. 
Step 3a) has a complexity of $ \myBigO{m^2} $, which only accounts for computing $ (\breve{\mathbf{H}}^{\mathrm{H}}\breve{\mathbf{H}})^{-1} $
in the null space. The other matrix multiplication will be jointly evaluated with Step 3b-i). Based on (\ref{eq: b star next (M-1)}) and (\ref{eq: N bf B*}), the complexity of Step 3b-i) is given by $ \myBigO{M+Mm+m^2+m} $, where $ m $ is also the cardinality of $ \mathcal{B}^{\star} $ in the $ m $-th iteration. 
According to Step 3b), Step 3b-i) is performed for up to $ (N-m) $ times in the $ m $-th $ (m\in [1,M-1]) $ iteration.
To sum up, the overall complexity of the beam selection is
\begin{align} \label{eq: complexity beam selection Alg 1}
	&\sum_{m=1}^{M-1} \myBigOsquare{(N-m)(M+Mm+m^2+m)} + \myBigO{m^2}\nonumber\\
	& + \myBigO{MN}
	\myApprOverset{N\gg M} \myBigO{M^3N}.
\end{align}
The complexity is unrelated to $ K $ now, since the beam selection in Algorithm \ref{alg: user first and then beams} is based on the $ M $ selected users.

\subsubsection{Algorithm \ref{alg: user and beams together}}
We only illustrate the case of $ K\gg M $ for brevity. In this case, the algorithm runs up to Step 4). Step 2) has a complexity of $ \myBigO{KN} $. Step 3a-i) can share the intermediate results of Step 2) and hence does not incur any extra complexity. Step 3a-ii) has a complexity of $ \myBigO{m^2} $, which again only accounts for computing $ (\breve{\mathbf{H}}^{\mathrm{H}}\breve{\mathbf{H}})^{-1} $
in the null space. Step 3a-iii) has a complexity of $ \myBigOsquare{(m+1)+(m+1)m+m^2+m}$, where $ (m+1) $ is the size of $ \mathcal{I} $ and $ m $ is the size of $ \mathcal{U}^{\star} $; see (\ref{eq: Pu}). As per Step 3a), Step 3a-ii) and Step 3a-iii) are performed $ (K-m) $
times for the $ m $-th $ (m\in[1,M-1]) $ iteration. In a nutshell, the overall complexity is 
\begin{align} \label{eq: complexity algorithm 2}
	& \sum_{m=1}^{M-1} \myBigOsquare{(K-m)((m+1)+(m+1)m+m^2+m+m^2)} \nonumber\\
	&+ \myBigO{KN} \myApprOverset{K\gg M} \myBigO{KM^3}.
\end{align}

\subsubsection{Algorithm \ref{alg: simul user and beams low complexity}}
Based on (\ref{eq: SIR}), computing $ \gamma_u^i $
has a complexity of $ \myBigOsquare{ (m+1) + m(m+1) + m} $. Note that the size of $ \tilde{\mathcal{I}}_u^i $ therein is $ (m+1) $, and the size of the matrix $ \check{\mathbf{H}} $ is $ m\times (m+1) $.
Since the matrix inversion in Step 3a-ii) is no longer necessary, the complexity of Step 3a-ii) and Step 3a-iii) becomes just $ \myBigOsquare{ (m+1) + m(m+1) + m} $. Similar to (\ref{eq: complexity algorithm 2}), the overall complexity is
\begin{align} \label{eq: complexity algorithm 3}
	& \sum_{m=1}^{M-1} \myBigOsquare{(K-m)((m+1) + m(m+1) + m)} \nonumber\\
	&+ \myBigO{KN} \myApprOverset{K\gg M} \myBigO{\frac{1}{3}KM^3}.
\end{align}
Strictly, the coefficient $ \frac{1}{3} $ should be removed under the big-O operation, but we keep it to show that Algorithm \ref{alg: simul user and beams low complexity} can indeed further reduce the complexity of Algorithm \ref{alg: user and beams together}.

{\textit{Though DPC has a higher complexity than ZF from the precoding perspective, the overall complexity of the communication transmitter, with the beam and user selections taken into account, can actually be reduced substantially by using DPC.} 
	This is because the monotonic evolution of the DPC sum rate against $ U $ and $ B $, as unveiled in Section \ref{sec: simplify selection problem}, enables us to simply take $ U= B=M $ and hence eliminate the necessity of enumerating all possible combinations of $ (U,B) $ with $ U=1,\cdots,M $ and $ B=1,\cdots,M $. In contrast, the state-of-the-art design based on ZF requires to enumerate all those combinations.

	{With $ M $ users and also $ M $ beams selected, the DPC precoder involves the QR decomposition of an $ M $-dimensional matrix; see (\ref{eq: P = QRD}). Therefore, the complexity of the DPC precoder is on the order of $ \myBigO{M^3} $. Jointly considering the complexity orders derived in (\ref{eq: complexity of user selection in Alg 1})-(\ref{eq: complexity algorithm 3}), we can assert that the complexity of the overall transmitting scheme proposed for the specific beamspace massive MIMO investigated in this work is proportional to $ M^3 $. In contrast, for the beam selection solely, most state-of-the-art designs, which employ ZF, already have a complexity order in proportion to $ N^3 $ \cite{Lens_beamSelection_XinyuGao2016,Lens_beamSelec_TCOM2015_Masouros,Lens_BeamSelect2stages_2019,Lens_BeamSelect_coplexityReduc2021_WCL}. 
		Moreover, the complexity of the state-of-the-art ZF-aided precoder developed for phase shifters-based hybrid massive MIMO arrays is also proportional to $ {N^3} $ \cite{massiveMIMO_precoding_survey2018}. Considering that $ N $ is generally on the order of hundreds while $ M $ is only on the order of dozens, our design becomes very appealing for the practical implementation of mmWave/THz massive MIMO communications.}	
}

\section{Simulation Results} \label{sec: simulation results}

\begin{table}[!t]\footnotesize
	\captionof{table}{Simulation Parameters}
	\vspace{-3mm}
	\begin{center}
		\begin{tabular}{m{0.05\linewidth}|m{0.38\linewidth}|m{0.42\linewidth}}
			\hline
			Var. &
			Description &
			Value
			\\
			\hline
			$ N $    
			&  antenna/beam number
			&	$ 256 $ 
			\\				
			\hline
			$ K $ 
			& total user number 
			& $ 40 $ 
			\\				
			\hline
			$ M $ 
			& RF chain number
			& $ 16 $
			\\				
			\hline
			$ \sigma_z^2 $ 
			& noise power; see (\ref{eq: S DPC})
			& $ 1 $ ($ 0 $ dB)
			\\				
			\hline
			$ \mP $ 
			& transmission power in Watt
			& $ [0,30] $ dB
			\\				
			\hline
			$ \mP/\sigma_z^2 $
			& SNR
			& $ [0,30] $ dB 
			\\				
			\hline
			-
			& channel path number per user
			& $ 3 $ (one LoS and two NLoS paths)
			\\				
			\hline
			- 
			& channel power
			& $ 0 $ dB for LoS; $ -10 $ dB for NLoS
			\\				
			\hline	
			-
			& AoD of all paths
			& uniform in $ [-90^{\circ},90^{\circ}] $
			\\				
			\hline

		\end{tabular}
		\vspace{-3mm}
	\end{center}
	\label{tab: simulation parameters}
\end{table}	

Simulation results are provided to validate the proposed designs, in comparison to both the performance bounds derived and to existing benchmark methods. 
Unless otherwise stated, the parameter settings of Table \ref{tab: simulation parameters} will be used. 
The settings in the table are popularly used in previous user and/or beam selection studies \cite{Lens_beamSelection_XinyuGao2016,Lens_JonitUserBeamSelection_2020,Lens_beamSelectionDPC_2018Pal,Lens_BeamSelect_coplexityReduc2021_WCL}. 
 
{As reviewed in Section \ref{sec: introduction}, most existing contributions in beamspace massive MIMO communications only aim for beam selection. To the best of our knowledge, the work in \cite{Lens_JonitUserBeamSelection_2020} is the only one that considers joint user and beam selection, and hence its is the most closely related piece of work. However, the solution in \cite{Lens_JonitUserBeamSelection_2020} is unsuitable as a benchmarker.
	One reason for this is that DPC is used in our work, while ZF is used in \cite{Lens_JonitUserBeamSelection_2020}. Since DPC is known to outperform ZF in terms of its sum rate \cite{book_mimoOFDMmatlab}, we cannot claim that the performance gain arises from our joint user and beam selection algorithms. 
	Moreover, as illustrated in Proposition \ref{pp: sum rate vs B U}, satisfying $ U=M $ and $ B=M $ is sufficient for maximizing the DPC sum rate, where $ U $ ($ B $) is the number of selected users (beams). By contrast, as shown in \cite{Lens_JonitUserBeamSelection_2020}, the ZF sum rate can be maximized when $ U<M $. As such, it would be unfair to compare our work to \cite{Lens_JonitUserBeamSelection_2020}.

Nevertheless, when the total number of users and RF chains is the same, i.e., $ K=M $, 
the user selection can be skipped in Algorithm \ref{alg: user first and then beams}, i.e., only the beam selection has to be performed. 
Recall that a greedy search is performed in Algorithm \ref{alg: user first and then beams} for the beam selection to maximize the DPC sum rate. This is also pursued by the designs in \cite{Lens_beamSelectionDPC_2018Pal,Lens_BeamSelect_coplexityReduc2021_WCL}. Thus, Algorithm \ref{alg: user first and then beams} essentially becomes indistinguishable from the designs in \cite{Lens_beamSelectionDPC_2018Pal,Lens_BeamSelect_coplexityReduc2021_WCL} in the case of $ K=M $. 
Furthermore, we note that the upper bounds, as derived in Section \ref{sec: performance complexity analysis}, 
provide a meaningful way of evaluating the performances of the proposed algorithms in both cases of $ K\gg M $ and $ K=M $.}

\begin{figure}[!t]
	\centering
	\includegraphics[width=85mm]{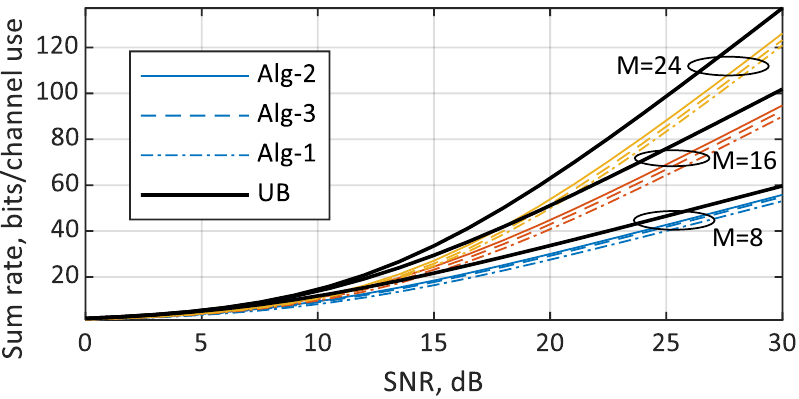}
	\caption{Sum rate versus SNR for the case of $ K>M $, where $ K=40 $ in (\ref{eq: max S}), UB stands for upper bound, as given in Corollary \ref{col: sum rate upper bound}.}
	\label{fig: sum rate vs snr KggM}
\end{figure}

Fig. \ref{fig: sum rate vs snr KggM} plots the sum rate versus SNR, where $ K=40 $ is higher than the three $ M $ values employed. We see that Algorithm \ref{alg: user and beams together} achieves the best performance (closest to the upper bound derived in Corollary \ref{col: sum rate upper bound}), Algorithm \ref{alg: simul user and beams low complexity} is the second, and Algorithm \ref{alg: user first and then beams} is the third. This first consolidates our motivation of developing Algorithm \ref{alg: user and beams together}, i.e., avoiding extra interference of using the full channel information as in Algorithm \ref{alg: user and beams together}. This also demonstrates the success of Algorithm \ref{alg: user and beams together} in so doing. Observed from Fig. \ref{fig: sum rate vs snr KggM}, Algorithm \ref{alg: simul user and beams low complexity} achieves a similar  performance to that of Algorithm \ref{alg: user and beams together}. This validates the effectiveness of using the SIR to replace the null space projection, as developed in Section \ref{subsec: sir based user and beam selection}. Given the small performance gap and the lower complexity quantified in Section \ref{subsec: compelxity}, Algorithm \ref{alg: simul user and beams low complexity} may be viewed as being more attractive than Algorithm \ref{alg: user and beams together}.

\begin{figure}[!t]
	\centering
	\includegraphics[width=85mm]{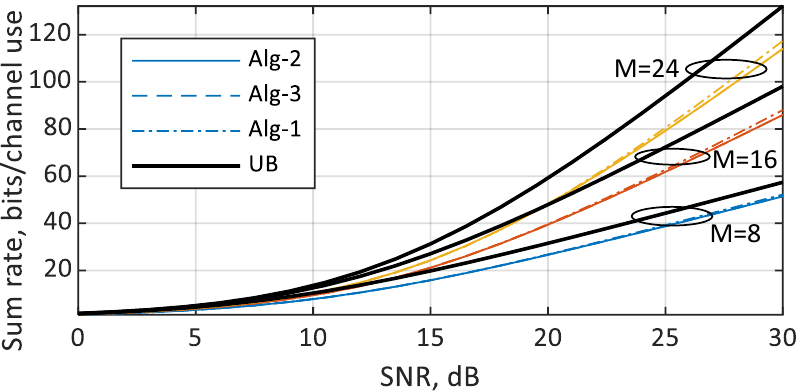}
	\caption{Sum rate versus SNR for the case of $ K=M $ in (\ref{eq: max S}), where UB stands for upper bound, as given in Corollary \ref{col: sum rate upper bound}. {Given $ K=M $, the user selection in Algorithm \ref{alg: user first and then beams} can be skipped, and hence only the beam selection is performed. Thus, Algorithm \ref{alg: user first and then beams} essentially corresponds to the benchmark methods \cite{Lens_beamSelectionDPC_2018Pal,Lens_BeamSelect_coplexityReduc2021_WCL}.}}
	\label{fig: sum rate vs snr K=M}
\end{figure}

Fig. \ref{fig: sum rate vs snr K=M} observes the sum rate versus SNR under $ K=M $, where no user selection is necessary. As stated earlier, this is a sole beam selection scenario, and Algorithm \ref{alg: user and beams together} may be regarded as the benchmark method \cite{Lens_beamSelectionDPC_2018Pal}. As indicated by Fig. \ref{fig: sum rate vs snr K=M}, the three algorithms achieve similar sum rate performance, but Algorithm \ref{alg: user first and then beams} is slightly better in the high-SNR region. This validates the result in Corollary \ref{col: alg 2 better K=M}. This also confirms the analysis in Remark \ref{rmk: K=M}. Moreover, 
our proposed algorithms have much lower complexity than the prior art \cite{Lens_beamSelectionDPC_2018Pal}. Jointly observing Figs. \ref{fig: sum rate vs snr KggM} and \ref{fig: sum rate vs snr K=M}, we can infer that the sum rate is closer to the upper bound in the case of $ K>M $. Intuitively, this is because the higher the number of users, the more likely we can select a small subset of users having better angular orthogonality; see also Proposition \ref{pp: M threashold} and Corollary \ref{col: channel orthogonal asymp}.

\begin{figure}[!t]
	\centering
	\includegraphics[width=85mm]{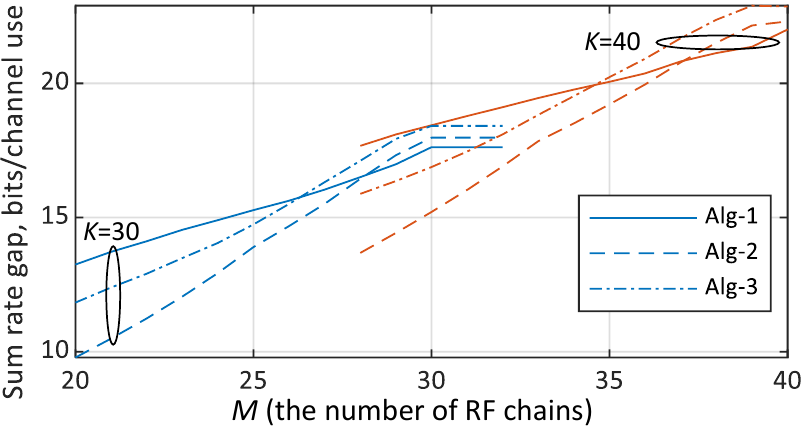}
	\caption{Sum rate against $ M $ in (\ref{eq: max S}), where the SNR is $ 28 $ dB. The gap is the difference between the upper bound and the sum rate achieved by an algorithm.}
	\label{fig: sum rate vs M}
\end{figure}

Fig. \ref{fig: sum rate vs M} illustrates the sum rate gap versus the number of RF chains $ (M) $, where the gap is the difference between the upper bound, as given in Corollary \ref{col: sum rate upper bound}, and the sum rate achieved by one of the algorithms. So the gap can be viewed as the sum rate loss. 
We observe that before $ M $ increases to a certain value, Algorithm \ref{alg: user and beams together} and \ref{alg: simul user and beams low complexity} have lower sum rate loss than Algorithm \ref{alg: user first and then beams}. This again validates our analysis in Remark \ref{rmk: K gg M}. We also see that, in all cases, Algorithm \ref{alg: user and beams together} has lower sum rate loss than Algorithm \ref{alg: simul user and beams low complexity}. This is to be expected, since Algorithm \ref{alg: simul user and beams low complexity} is based on an approximate orthogonality, while Algorithm \ref{alg: user and beams together} builds on the orthogonal null space projection. From Fig. \ref{fig: sum rate vs M} we can also see an interesting change in the performance gain of Algorithm \ref{alg: user and beams together}/\ref{alg: simul user and beams low complexity} over that of \ref{alg: user first and then beams}. There is an intercept point, beyond which the performance gain turns from positive to negative. This corresponds to the threshold of $ \bar{M} $, as derived in Proposition \ref{pp: M threashold}. However, the positive gain has a much higher value.

\begin{figure}[!t]
	\centering
	\includegraphics[width=85mm]{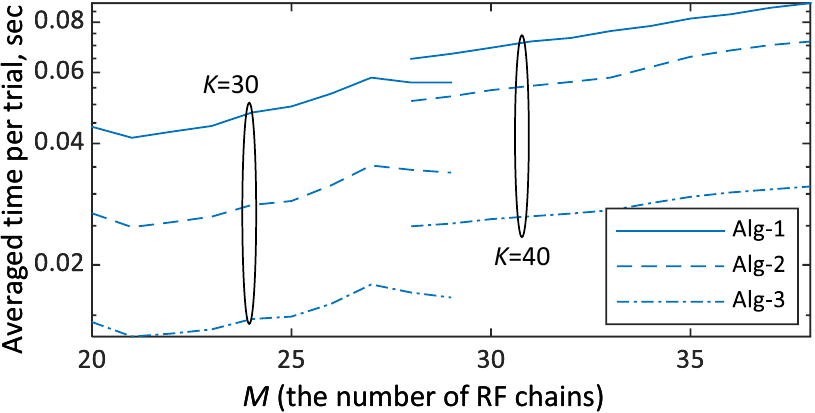}
	\caption{The average time of each algorithm per run, where $ 5\times 10^3 $ independent trials are performed to get the result.}
	\label{fig: runtime}
\end{figure}

{Fig. \ref{fig: runtime} compares the average running time of the three algorithms, where different numbers of RF chains $ (M) $ and users $ (K) $ are considered. The other parameters are set the same as those used for Fig. \ref{fig: sum rate vs M}. 
	In accordance with the complexity analysis in Section \ref{subsec: compelxity}, Algorithm 1 has the longest running time, while Algorithm 2 has the shortest. Moreover, as $ M $ increases, the running time of all algorithms exhibit increasing trends, which complies with the complexity orders derived in (\ref{eq: complexity of user selection in Alg 1})-(\ref{eq: complexity algorithm 3}). In addition, as $ K $ increases, the complexity gap between Algorithms \ref{alg: user and beams together} and \ref{alg: simul user and beams low complexity} increases, while that between Algorithms \ref{alg: user first and then beams} and \ref{alg: user and beams together} decreases. This has also been predicted by our analysis. In particular, we see from (\ref{eq: complexity beam selection Alg 1}) and (\ref{eq: complexity algorithm 2}) that the difference in their complexity is related to $ (N-K) $. Hence, given a fixed $ N $, a higher $ K $ leads to a smaller difference in the running times of Algorithms \ref{alg: user first and then beams} and \ref{alg: user and beams together}, as observed in Fig. \ref{fig: runtime}. Moreover, as indicated by (\ref{eq: complexity algorithm 2}) and (\ref{eq: complexity algorithm 3}), the difference of the running time between Algorithms \ref{alg: user and beams together} and \ref{alg: simul user and beams low complexity} increases with $ K $, hence leading to the trends seen in the figure.
}

\begin{figure}[!t]
	\centering
	\includegraphics[width=85mm]{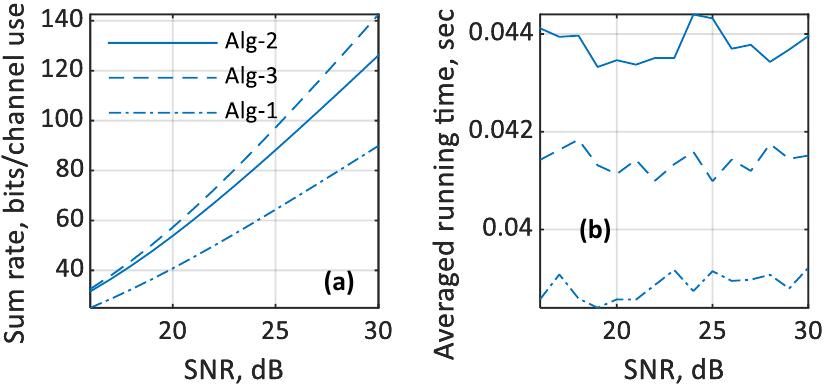}
	\caption{The average time of each algorithm per run, where $ 5\times 10^3 $ independent trials are performed to get the result.}
	\label{fig: sum rate vs snr diff M for algs}
\end{figure}

{Fig. \ref{fig: sum rate vs snr diff M for algs} further compares the performances of the three algorithms under similar running time. According to our analysis in Section \ref{sec: performance complexity analysis}, a smaller $ M $ corresponds to the lower complexity. Therefore, to arrange for the algorithms to have similar complexity, we set $ K=40 $ for the algorithms but set $ M=16,24 $ and $ 32 $ for Algorithms \ref{alg: user first and then beams}, \ref{alg: user and beams together} and \ref{alg: simul user and beams low complexity}, respectively. In this spirit, Fig. \ref{fig: sum rate vs snr diff M for algs}(b) confirms the similar complexity of the three algorithms, while Fig. \ref{fig: sum rate vs snr diff M for algs}(a) shows that Algorithms \ref{alg: user first and then beams}, \ref{alg: user and beams together} and \ref{alg: simul user and beams low complexity} exhibit worst, second and best sum rate performances, respectively. This is consistent with the DPC sum rate trend unveiled in Section \ref{sec: simplify selection problem}; specifically, the sum rate increases with the number of selected users under sufficient power budget. It is interesting to observe that the sum rate gap between Algorithms 1 and 2 is much larger than that between Algorithms 2 and 3. This is attributable to the increasingly degraded channel orthogonality, as $ M $ increases.}

\begin{figure}[!t]
	\centering
	\includegraphics[width=85mm]{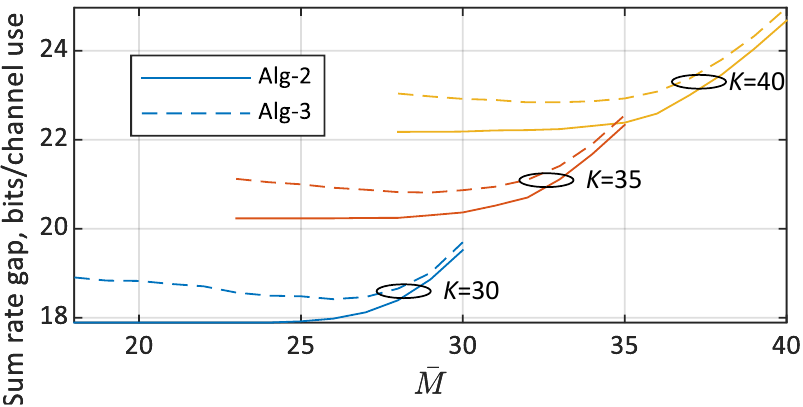}
	\caption{Sum rate gap versus $ \bar{M} $ in (\ref{eq: M threshold}), where $ \mP=28 $ dB; see (\ref{eq: max S simplified}), and $ M=30,35 $ and $ 40 $ for the same values of $ K $.}
	\label{fig: sum rate vs Mbar}
\end{figure}

Fig. \ref{fig: sum rate vs Mbar} illustrates the sum rate gap versus $ \bar{M} $ which is an intermediate variable used in Step 4) of Algorithm \ref{alg: user and beams together}.
We use Proposition \ref{pp: M threashold}, the upper limit of $ \bar{M}$ and $\myFloor{\frac{KN}{K+N}}=26,30 $ and $ 34 $ for $ K=30 $, $ 35 $ and $ 40 $, respectively. (Note that the corresponding upper limits of $ \bar{M} $ are used for the previous figures.) Observed from Fig. \ref{fig: sum rate vs Mbar}, the sum rate remains near-constant when $ \bar{M} $ is lower than the upper limit. However, the sum rate gap increases, as $ \bar{M} $ exceeds the upper limit. These results further confirm the validity of the analysis in Proposition \ref{pp: M threashold} and the precision of the upper limit of $ \bar{M} $ derived therein. Moreover, we emphasize that the upper limit should be used whenever possible, since it minimizes the complexity of Algorithm \ref{alg: user and beams together} if Step 5) does not need to be performed.

\begin{figure}[!t]
	\centering
	\includegraphics[width=85mm]{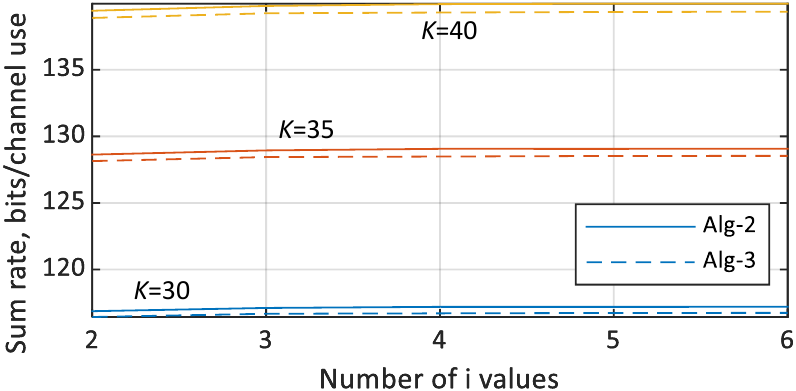}
	\caption{Sum rate versus the number of $ i $ values used in Step 5a) of Algorithm \ref{alg: user and beams together}, where $ \mP=28 $ dB, and $ M=30,35 $ and $ 40 $ for the same values of $ K $.}
	\label{fig: sum rate vs i No.}
\end{figure}

Fig. \ref{fig: sum rate vs i No.} quantifies the sum rate performance versus the numbers of $ i $ values used in Step 5a) of Algorithm \ref{alg: user and beams together}. 
Recall that $ i $ is the index offset introduced to help us avoid selecting the same beam as previously selected; see Section \ref{subsec: simultaneous user and beam selection}. From Fig. \ref{fig: sum rate vs i No.}, we can only see a slight improvement as the number of $ i $ values increases from two to three. This confirms the validity of only taking three values of $ i $, i.e., $ 0 $ and $ \pm 1 $ in Algorithm \ref{alg: user and beams together}. We underline that ``three'' is not a guess, the rationale has been illustrated in Section \ref{subsec: simultaneous user and beam selection}.

\section{Conclusions} \label{sec: conclusion}
The joint user and beam selection problem of the beamspace mmWave/THz massive MIMO downlink was studied. 
Motivated by its appealingly simple sum rate expression, we harnessed DPC to the joint selection problem for the first time in the open literature. We unveiled that the DPC sum rate monotonically increases with the number of selected users and it is a non-decreasing function of the number of selected beams. These features allow us to substantially simplify the user and beam selection problem.  
We then developed three algorithms for solving the simplified problem. The first one sequentially selects users and then beams, while the second one simultaneously selects both for reducing the extra inter-user interference encountered in the first. While the null space projection is used as a performance metric, the third algorithm further reduces the complexity by replacing the null projection by the SIR. 
The performance-vs-complexity analysis was carried out for the three algorithms. Our simulation results validated the efficiency of the proposed designs.

{Several exciting future research problems are suggested next. First, the analysis and the proposed algorithms may be extended to user grouping and beam selection, another critical problem in beamspace massive MIMO  \cite{Lens_BeamUserGroup}. Moreover, the applicability and extension of the proposed methods to more advanced beamspace MIMO scenarios/settings, such as wideband and phase shifters-enabled beam selection \cite{Lens_BeamSelectWideband2019_TSP,Lens_BeamSelectSLNR_2022}, are worth further investigating.} In addition, the proposed designs may be extended by further considering the beam selection on the user side \cite{BeamSelection_userSide_TCOM2015}. 
Finally, the rate-fairness of users suffering from low SNR may be improved by relying on geometric mean rate based {optimization \cite{GM_OPt_rate_fairness}.}

\appendix
\subsection{Proof of Lemma \ref{lm: prod(r_u) vs B}} \label{app: proof of lemma prod(r_u) vs B}
Since $ \mathbf{R} $ is an upper triangular matrix, we have
$ {\Pi_{u=0}^{U-1}r_u} = { \myDet{\mathbf{R}} } $. Multiplying $ \mathbf{R} $ by an orthogonal matrix, e.g., $ \mathbf{Q} $ in (\ref{eq: U H B = QR}), does not change its determinant. Thus, we have 
\begin{align}
	{\Pi_{u=0}^{U-1}r_u} = { \myDet{\mathbf{QR}} } &= \Pi_{u=0}^{U-1}\sigma_u \nonumber\\
	&= \sqrt{ \Pi_{u=0}^{U-1}\eta_u } = \sqrt{\myDet{\mathbf{\Gamma}}}, \nonumber
\end{align}
where $ \sigma_u $ is the $ u $-th singular value of $ (\mathbf{U} \tilde{\mathbf{H}} \mathbf{B}) $, $ \eta_u $ is the eigenvalue of $ \mathbf{\Gamma} $, and 
$\mathbf{\Gamma} = (\mathbf{U} \tilde{\mathbf{H}} \mathbf{B})(\mathbf{U} \tilde{\mathbf{H}} \mathbf{B})^{\mathrm{H}} $. As a Hermitian matrix, $ \myDet{\mathbf{\Gamma}} $ is non-decreasing with $ B $; so is $ {\Pi_{u=0}^{U-1}r_u}=\sqrt{\myDet{\mathbf{\Gamma}}}  $.

\subsection{Proof of Proposition \ref{pp: M threashold}}
\label{app: prood of propostion on M threshold}
Assume that $ m(\in [1,M-1]) $ users have been selected, and their strongest beams have different indexes as collected by $ \mathcal{B} $.  
Let us now
consider the probability $ \mathcal{P} $ of the event $ \mathcal{V} $: the index of the strongest beam of the next selected user is in 
$ \mathcal{B} $, where $ \mathcal{P} $ is the product of the probabilities of two independent events: 
\begin{enumerate}
	\item The first one is that one out of $ (K-m) $ remaining users is selected next. Its probability is $ \mathcal{P}_1=\frac{1}{K-m} $, since these users have equal chance of being selected.
	
	\item The second event is that the selected user shares the strongest beam with any one of the $ m $ pre-selected users. 
	Its probability is $ \mathcal{P}_2 = \frac{m}{N} $, as the strongest beam of each user can be one of the $ N $ beams with the same probability.
\end{enumerate}
Using the two probabilities, we obtain $ \mathcal{P}=\mathcal{P}_1\mathcal{P}_2 =\frac{m}{(K-m)N} $.

Given $ K $ users and the probability $ \mathcal{P} $, $ {KP}< 1 $ implies that less than one, out of $ (K-m) $ remaining users, will share the strongest beam with one of the $ m $ selected users; in other words, the event $ \mathcal{V} $ almost never happens. Solving the following problem leads to the upper limit of $ m $:
		\begin{align} \label{eq: M bar problem}
				\bar{M}=\myFloor{m^{\star}},~\mathrm{s.t.}~ { \frac{{m^{\star}}K}{(K-{m^{\star}})N}} = 1,
			\end{align}		
	where $ \myFloor{\cdot} $ rounds the enclosed number towards negative infinity.
Provided that $ \bar{M}\ge M $ (the number of RF chains), we can state almost for sure that $ \mathcal{V} $ cannot happen, or otherwise the event $ \mathcal{E} $ given in the statement of Proposition \ref{pp: M threashold} always happens.

\bibliographystyle{IEEEtran}
\bibliography{IEEEabrv,./bib_JCAS.bib}

\begin{thebibliography}{10}
\providecommand{\url}[1]{#1}
\csname url@samestyle\endcsname
\providecommand{\newblock}{\relax}
\providecommand{\bibinfo}[2]{#2}
\providecommand{\BIBentrySTDinterwordspacing}{\spaceskip=0pt\relax}
\providecommand{\BIBentryALTinterwordstretchfactor}{4}
\providecommand{\BIBentryALTinterwordspacing}{\spaceskip=\fontdimen2\font plus
\BIBentryALTinterwordstretchfactor\fontdimen3\font minus
  \fontdimen4\font\relax}
\providecommand{\BIBforeignlanguage}[2]{{%
\expandafter\ifx\csname l@#1\endcsname\relax
\typeout{** WARNING: IEEEtran.bst: No hyphenation pattern has been}%
\typeout{** loaded for the language `#1'. Using the pattern for}%
\typeout{** the default language instead.}%
\else
\language=\csname l@#1\endcsname
\fi
#2}}
\providecommand{\BIBdecl}{\relax}
\BIBdecl

\bibitem{6G_vision_2020network}
W.~Saad, M.~Bennis, and M.~Chen, ``A vision of {6G} wireless systems:
  Applications, trends, technologies, and open research problems,'' \emph{IEEE
  Network}, vol.~34, no.~3, pp. 134--142, 2020.

\bibitem{6G_XiaohuYou2021towards}
X.~You, C.-X. Wang, J.~Huang, X.~Gao, Z.~Zhang, M.~Wang, Y.~Huang, C.~Zhang,
  Y.~Jiang, J.~Wang \emph{et~al.}, ``Towards {6G} wireless communication
  networks: Vision, enabling technologies, and new paradigm shifts,''
  \emph{Science China Information Sciences}, vol.~64, no.~1, pp. 1--74, 2021.

\bibitem{6G_enable2020Access}
L.~Bariah, L.~Mohjazi, S.~Muhaidat, P.~C. Sofotasios, G.~K. Kurt,
  H.~Yanikomeroglu, and O.~A. Dobre, ``A prospective look: Key enabling
  technologies, applications and open research topics in {6G} networks,''
  \emph{IEEE Access}, vol.~8, pp. 174\,792--174\,820, 2020.

\bibitem{Lajos_mmWaveCom_survey2018}
I.~A. Hemadeh, K.~Satyanarayana, M.~El-Hajjar, and L.~Hanzo, ``Millimeter-wave
  communications: Physical channel models, design considerations, antenna
  constructions, and link-budget,'' \emph{IEEE Commun. Surv. Tutor.}, vol.~20,
  no.~2, pp. 870--913, 2018.

\bibitem{Lens_lens5GlowPower_2018Mag}
X.~Gao, L.~Dai, and A.~M. Sayeed, ``Low {RF}-complexity technologies to enable
  millimeter-wave {MIMO} with large antenna array for {5G} wireless
  communications,'' \emph{IEEE Commun. Mag.}, vol.~56, no.~4, pp. 211--217,
  2018.

\bibitem{KaiWu_WIPT_LAA}
K.~Wu, W.~Ni, T.~Su, R.~P. Liu, and Y.~J. Guo, ``Efficient {Angle-of-Arrival}
  estimation of lens antenna arrays for wireless information and power
  transfer,'' \emph{IEEE J. Sel. Areas Commun.}, vol.~37, no.~1, pp. 116--130,
  Jan 2019.

\bibitem{Lens_PathDivision_TCOM2016}
Y.~Zeng and R.~Zhang, ``Millimeter wave {MIMO} with lens antenna array: A new
  path division multiplexing paradigm,'' \emph{IEEE Trans. Commun.}, vol.~64,
  no.~4, pp. 1557--1571, 2016.

\bibitem{Jay_Butler6G_overview}
Y.~J. Guo, M.~Ansari, and N.~J.~G. Fonseca, ``Circuit type multiple beamforming
  networks for antenna arrays in {{5G}} and {6G} terrestrial and
  non-terrestrial networks,'' \emph{IEEE J. Microw.}, vol.~1, no.~3, pp.
  704--722, 2021.

\bibitem{Jay_lens6G_overview}
Y.~J. Guo, M.~Ansari, R.~W. Ziolkowski, and N.~J.~G. Fonseca, ``Quasi-optical
  multi-beam antenna technologies for {B5G} and {6G} mmwave and {THz} networks:
  A review,'' \emph{IEEE Open J. Antennas Propag.}, vol.~2, pp. 807--830, 2021.

\bibitem{Beamspace_sayeed2013beamspaceMIMO}
A.~Sayeed and J.~Brady, ``Beamspace {{MIMO}} for high-dimensional multiuser
  communication at millimeter-wave frequencies,'' in \emph{2013 IEEE
  GLOBECOM}.\hskip 1em plus 0.5em minus 0.4em\relax IEEE, 2013, pp. 3679--3684.

\bibitem{Lajos_BmSlctPrecdng_JSTSP2018}
R.~Guo, Y.~Cai, M.~Zhao, Q.~Shi, B.~Champagne, and L.~Hanzo, ``Joint design of
  beam selection and precoding matrices for mmwave {MU-MIMO} systems relying on
  lens antenna arrays,'' \emph{IEEE J. Sel. Top. Signal Process.}, vol.~12,
  no.~2, pp. 313--325, 2018.

\bibitem{Lajos_beamspaceWidebandChannelEstimation_TSP2019}
X.~Gao, L.~Dai, S.~Zhou, A.~M. Sayeed, and L.~Hanzo, ``Wideband beamspace
  channel estimation for millimeter-wave {MIMO} systems relying on lens antenna
  arrays,'' \emph{IEEE Trans. Signal Process.}, vol.~67, no.~18, pp.
  4809--4824, 2019.

\bibitem{Lens_beamspaceChannelEstimation_TCM2018}
J.~Yang, C.-K. Wen, S.~Jin, and F.~Gao, ``Beamspace channel estimation in
  mmwave systems via cosparse image reconstruction technique,'' \emph{IEEE
  Trans. Commun.}, vol.~66, no.~10, pp. 4767--4782, 2018.

\bibitem{Lens_beamSelection_XinyuGao2016}
X.~Gao, L.~Dai, Z.~Chen, Z.~Wang, and Z.~Zhang, ``Near-optimal beam selection
  for beamspace mmwave massive {{MIMO}} systems,'' \emph{IEEE Commun. Lett.},
  vol.~20, no.~5, pp. 1054--1057, 2016.

\bibitem{Lens_beamSelec_TCOM2015_Masouros}
P.~V. Amadori and C.~Masouros, ``Low {RF}-complexity millimeter-wave
  beamspace-{{MIMO}} systems by beam selection,'' \emph{IEEE Trans. Commun.},
  vol.~63, no.~6, pp. 2212--2223, 2015.

\bibitem{Lens_BeamSelect2stages_2019}
H.~Yu, W.~Qu, Y.~Fu, C.~Jiang, and Y.~Zhao, ``A novel two-stage beam selection
  algorithm in mmwave hybrid beamforming system,'' \emph{IEEE Commun. Lett.},
  vol.~23, no.~6, pp. 1089--1092, 2019.

\bibitem{Lens_beamSelectionDPC_2018Pal}
R.~Pal, K.~V. Srinivas, and A.~K. Chaitanya, ``A beam selection algorithm for
  millimeter-wave multi-user {{MIMO}} systems,'' \emph{IEEE Commun. Lett.},
  vol.~22, no.~4, pp. 852--855, 2018.

\bibitem{Lens_BeamSelect_coplexityReduc2021_WCL}
Q.~Zhang, X.~Li, B.-Y. Wu, L.~Cheng, and Y.~Gao, ``On the complexity reduction
  of beam selection algorithms for beamspace {{MIMO}} systems,'' \emph{IEEE
  Wireless Commun. Lett.}, vol.~10, no.~7, pp. 1439--1443, 2021.

\bibitem{Lens_JonitUserBeamSelection_2020}
Z.~Cheng, Z.~Wei, and H.~Yang, ``Low-complexity joint user and beam selection
  for beamspace mmwave {{MIMO}} systems,'' \emph{IEEE Commun. Lett.}, vol.~24,
  no.~9, pp. 2065--2069, 2020.

\bibitem{book_ahmadi2019_5G}
S.~Ahmadi, \emph{{{5G}} {NR}: Architecture, Technology, Implementation, and
  Operation of {3GPP} New Radio Standards}.\hskip 1em plus 0.5em minus
  0.4em\relax Academic Press, 2019.

\bibitem{Kai_Expeditious2019TWC}
K.~{Wu}, W.~{Ni}, T.~{Su}, R.~P. {Liu}, and Y.~J. {Guo}, ``Expeditious
  estimation of angle-of-arrival for hybrid butler matrix arrays,'' \emph{IEEE
  Trans. Wireless Commun.}, vol.~18, no.~4, pp. 2170--2185, April 2019.

\bibitem{Lens_xinyuGao_compressiveChannelEstimation}
X.~Gao, L.~Dai, S.~Han, C.-L. I, and X.~Wang, ``Reliable beamspace channel
  estimation for millimeter-wave massive {{MIMO}} systems with lens antenna
  array,'' \emph{IEEE Trans. Wireless Commun.}, vol.~16, no.~9, pp. 6010--6021,
  2017.

\bibitem{Beamspace_gao2016fast_THz_channeltracking}
X.~Gao, L.~Dai, Y.~Zhang, T.~Xie, X.~Dai, and Z.~Wang, ``Fast channel tracking
  for thz beamspace massive {{MIMO}} systems,'' \emph{IEEE Trans. Vehic.
  Techn.}, vol.~66, no.~7, pp. 5689--5696, 2016.

\bibitem{book_mimoOFDMmatlab}
Y.~S. Cho, J.~Kim, W.~Y. Yang, and C.~G. Kang, \emph{{{MIMO}-OFDM} wireless
  communications with {MATLAB}}.\hskip 1em plus 0.5em minus 0.4em\relax John
  Wiley \& Sons, 2010.

\bibitem{UserSelection_ZF_Dimic2005TSP}
G.~Dimic and N.~Sidiropoulos, ``On downlink beamforming with greedy user
  selection: performance analysis and a simple new algorithm,'' \emph{IEEE
  Trans. Signal Process.}, vol.~53, no.~10, pp. 3857--3868, 2005.

\bibitem{massiveMIMO_precoding_survey2018}
N.~Fatema, G.~Hua, Y.~Xiang, D.~Peng, and I.~Natgunanathan, ``Massive mimo
  linear precoding: A survey,'' \emph{IEEE Systems Journal}, vol.~12, no.~4,
  pp. 3920--3931, 2018.

\bibitem{Lens_BeamUserGroup}
J.~Sun, M.~Jia, Q.~Guo, and X.~Gu, ``Joint user grouping and beam selection for
  beamspace mmwave multi-user {{MIMO}} system,'' \emph{IEEE Commun. Lett.},
  vol.~26, no.~5, pp. 1170--1174, 2022.

\bibitem{Lens_BeamSelectWideband2019_TSP}
W.~Shen, X.~Bu, X.~Gao, C.~Xing, and L.~Hanzo, ``Beamspace precoding and beam
  selection for wideband millimeter-wave {{MIMO}} relying on lens antenna
  arrays,'' \emph{IEEE Trans. Signal Process.}, vol.~67, no.~24, pp.
  6301--6313, 2019.

\bibitem{Lens_BeamSelectSLNR_2022}
Z.~Cheng, Z.~Wei, H.~Li, and H.~Yang, ``{SLNR}-based beamspace precoding and
  beam selection for wideband mmwave massive {{MIMO}},'' \emph{IEEE Commun.
  Lett.}, vol.~26, no.~2, pp. 478--482, 2022.

\bibitem{BeamSelection_userSide_TCOM2015}
J.~Choi, ``Beam selection in mm-wave multiuser {MIMO} systems using compressive
  sensing,'' \emph{IEEE Trans. Commun.}, vol.~63, no.~8, pp. 2936--2947, 2015.

\bibitem{GM_OPt_rate_fairness}
H.~Yu, H.~D. Tuan, E.~Dutkiewicz, H.~V. Poor, and L.~Hanzo, ``Maximizing the
  geometric mean of user-rates to improve rate-fairness: Proper vs. improper
  gaussian signaling,'' \emph{IEEE Trans. Wirel. Commun.}, vol.~21, no.~1, pp.
  295--309, 2022.

\end{thebibliography}

\end{document}